\documentclass[
  aps,
  pra,
  reprint,
  amsmath,
  amssymb,
  superscriptaddress,
  nofootinbib,
  longbibliography
]{revtex4-2}

\usepackage[T1]{fontenc}
\usepackage{lmodern}
\usepackage{microtype}
\usepackage{mathtools,bm,amsthm}
\usepackage{booktabs,multirow}
\usepackage{graphicx}
\usepackage{xcolor}
\usepackage{enumitem}
\usepackage{hyperref}
\usepackage[nameinlink,noabbrev]{cleveref}

\definecolor{ink}{HTML}{172033}
\definecolor{flowblue}{HTML}{2457A6}
\definecolor{phasegreen}{HTML}{256D48}
\definecolor{failred}{HTML}{9D2A2A}
\hypersetup{
  colorlinks=true,
  linkcolor=flowblue,
  citecolor=phasegreen,
  urlcolor=flowblue,
  pdftitle={Structured Hamiltonian Learning for Multiqubit Conditional Phase Gates},
  pdfauthor={Xiu-Hao Deng},
  pdfsubject={Target-relative commuting error generators, gate-equivalence specifications, Ramsey-Walsh learning, and resource bounds},
  pdfkeywords={Hamiltonian learning, multiqubit conditional phase gate, QCVV, error generator, Ramsey measurement, Walsh transform, generalized controlled phase, sample complexity, calibration}
}

\setlist{nosep,leftmargin=1.35em}
\allowdisplaybreaks
\newtheorem{theorem}{Theorem}[section]
\newtheorem{proposition}[theorem]{Proposition}
\newtheorem{corollary}[theorem]{Corollary}

\theoremstyle{definition}

\theoremstyle{remark}

\newcommand{\Tr}{\operatorname{Tr}}
\newcommand{\rank}{\operatorname{rank}}
\newcommand{\diag}{\operatorname{diag}}
\newcommand{\Var}{\operatorname{Var}}
\newcommand{\Cov}{\operatorname{Cov}}
\newcommand{\Id}{\mathbb I}
\newcommand{\E}{\mathbb E}
\newcommand{\ket}[1]{\lvert #1\rangle}

\newcommand{\proj}[1]{\lvert #1\rangle\!\langle #1\rvert}

\newcommand{\e}{\mathrm e}
\newcommand{\wrap}{\operatorname{wrap}_{2\pi}}
\newcommand{\cG}{\mathcal G}
\newcommand{\cA}{\mathcal A}
\newcommand{\Mob}{\mathcal M}

\begin{document}

\title{Structured Hamiltonian Learning for Multiqubit Conditional Phase Gates}

\author{Xiu-Hao Deng}
\email{dengxiuhao@iqasz.cn}
\affiliation{International Quantum Academy, Shenzhen 518048, China}
\affiliation{Shenzhen Branch, Hefei National Laboratory, Shenzhen 518048, China}
\date{\today}

\begin{abstract}
Accurately measuring an intended conditional phase does not by itself establish
that a multiqubit phase gate belongs to its declared generalized-gate equivalence
class.  We formulate this diagnosis as structured Hamiltonian learning of the
diagonal successful block in a known computational basis.  Gate/reference Ramsey
measurements on a connected graph reconstruct a target-relative eigenphase map,
and a Walsh transform yields the stroboscopic error generator of one gate cycle,
$H_{\rm err}=\sum_SJ_SZ_S$, within a selected logarithm branch.  Because these real
coefficients are branch dependent, they cannot by themselves decide global gate
equivalence.  For the equivalence group generated by diagonal Pauli-$Z$ terms of
weight at most $m$, we derive an exact branch-independent criterion based on
multiplicative Boolean--M\"obius invariants.  This criterion permits arbitrary
allowed phase dressing, including large phases that generate spurious high-weight
coefficients in a principal logarithm.  At finite shot counts, prespecified
tolerances and a simultaneous confidence value convert the criterion into a
generalized-gate \textsc{pass}/\textsc{fail}/\textsc{unresolved} statistical
decision, separate from model rejection.

For shared-control $C^mZ^{\otimes n}$ gates, a circular target-face contrast---an
alternating Boolean finite difference modulo $2\pi$---isolates each intended
conditional phase.  A support theorem nevertheless exhibits forbidden interactions
that are invisible to every designated target face when multiple targets or external
spectators are present.  We derive a Cram\'er--Rao bound for an arbitrary integer
phase contrast and recover the $4^m$ attempted-circuit scaling of a uniform
$(m+1)$-body face as a special case.  Visibility loss, heralded survival, and readout
confusion reduce the Fisher information; correlated phase noise enters through the
face-sign covariance; systematic bias and violations of the diagonal model trigger
model rejection rather than an inflated error bar.  Seeded multinomial-count
simulations implement the $4M$-setting multiplexed acquisition end to end.  For
$M=3,4,5$ and a common attempted-circuit budget, its mean phase-map RMSE is
$0.689$--$0.794$ times that of a separately compiled statewise scan.  In a
multi-target blind-interaction test, the complete global screening audit rejects all 300
seeded replicates, whereas the target-face rejection rate is $1/300=0.003$.
These results estimate the stroboscopic error generator of a diagonal successful
block under a specified acquisition model; they do not constitute unknown-basis
many-body Hamiltonian identification, arbitrary-channel tomography, or hardware
certification.
\end{abstract}

\maketitle

\section{Introduction}
\label{sec:introduction}

Hamiltonian learning uses controlled observations to infer a structured dynamical
generator.  Bayesian, quantum-assisted, and experimental studies have established
the system-identification viewpoint
\cite{granade2012robust,wiebe2014quantum,wang2017experimental}.  Locality, sparsity,
access to real-time evolution, and prior knowledge can change both identifiability and resource scaling
\cite{bairey2019local,haah2024learning,gu2024practical,hu2025ansatzfree}.
For a processor, the learned generator can also serve as a diagnostic or feedback
coordinate: effective Floquet generators characterize stroboscopic blocks
\cite{pastori2022floquet}, active query selection has been demonstrated for a
two-qubit cross-resonance Hamiltonian \cite{dutt2023active}, and structured dynamical
models have been learned on superconducting devices \cite{hangleiter2024robust}.
Those settings motivate a narrower question: what is the correct learning object
when the operation itself is a multiqubit conditional phase gate?

The answer is unusually transparent but requires two qualifications.  If the successful
computational block is diagonal in a known basis, its coherent error has commuting
Pauli-$Z$ coordinates within that block.  This does not mean that arbitrary wrapped phases specify a
unique Hamiltonian logarithm, nor that learning those coordinates reconstructs a
microscopic driven Hamiltonian.  A selected logarithm branch and explicit
model-validity checks are necessary.  Second, the meaning of ``correct gate''
depends on which diagonal phase
dressings are declared equivalent or compensable.  A standard $CCCZ$ matrix, a
$CCCZ$ followed by virtual single-qubit $Z$ frames, and a gate dressed by arbitrary
three-local diagonal phases are three different target specifications.  Comparing
them with one unqualified fidelity metric can turn an allowed frame into an apparent error or hide a
forbidden many-body phase.

\begin{figure*}[t]
\centering
\includegraphics[width=\textwidth]{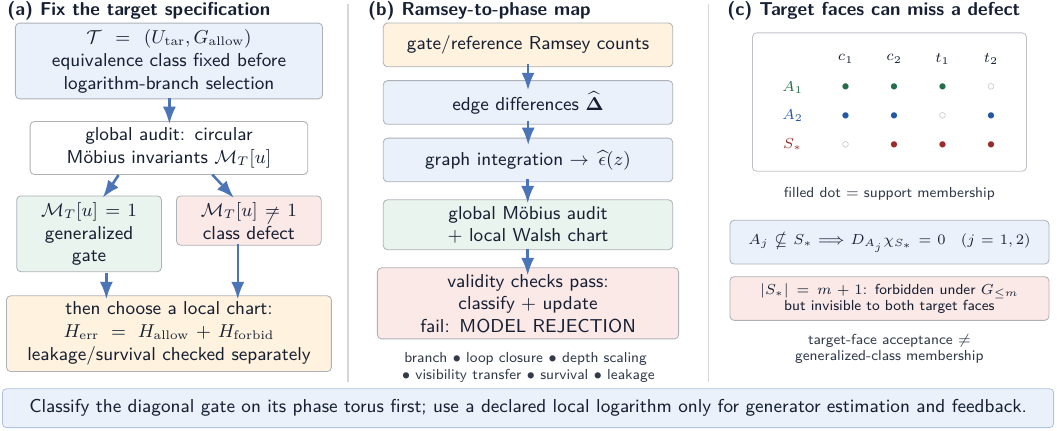}
\caption{Target-relative structured Hamiltonian learning.  (a) The allowed group is
fixed as part of the target contract.  Global multiplicative invariants decide class
membership; a selected logarithm supplies local generator coordinates for feedback.
(b) Gate/reference Ramsey counts give edge phases; graph integration gives the phase
field, followed by both the global M\"obius audit and the local Walsh transform.
Generator coordinates are reported for feedback only after the model-validity
checks pass.
(c) The support-membership rows make the incompleteness explicit: for two controls
and two targets, $S_*=\{c_2,t_1,t_2\}$ is forbidden under $\cG_{\le2}$ but contains
neither complete target face, so both designated face contrasts vanish.  A target-face
PASS is therefore necessary but not sufficient for quotient membership.}
\label{fig:pipeline}
\end{figure*}

We address both issues with a target-relative, quotient-aware formulation.  The
target specification, denoted the \emph{target contract}, is
$\mathcal T=(U_{\rm tar},\cG_{\rm allow})$ and is fixed before inspecting data.
Within the accepted logarithm branch, the target-relative phase map is represented
in the computational Pauli-$Z$ basis and decomposed as
$H_{\rm err}=H_{\rm allow}+H_{\rm forbid}$ for feedback.  Global equivalence is
tested separately on the phase torus, where multiplicative M\"obius invariants are
unchanged by independent $2\pi$ relabelings of basis-state phases.  Leakage and
nonuniform successful amplitudes remain outside both phase descriptions.  The
organizing logic is shown in \cref{fig:pipeline}.

Walsh functions are already a standard basis for diagonal operators and their
synthesis \cite{welch2014diagonal,hadfield2021boolean}; conditional Ramsey calibration,
repeated-gate phase amplification, and coherent error-generator feedback are
established for two-qubit phase gates \cite{xue2022spin,kimmel2015rpe,sung2021cz};
and complete or self-consistent process characterization has a broader scope
\cite{poyatos1997process,blumekohout2017gst}.  Direct fidelity estimation and
compressed device characterization instead target selected properties without full
process reconstruction \cite{flammia2011direct,dasilva2011practical}.

High-fidelity conditional-phase gates have been demonstrated in superconducting
circuits using leakage interference, tunable couplers, and deliberately nonadiabatic
trajectories \cite{rol2019fast,sung2021cz,negirneac2021cz}.  Neutral-atom arrays
provide parallel controlled-phase gates and native multiqubit entangling operations
\cite{levine2019parallel,evered2023highfidelity}.  These experiments motivate the
gate family, but they do not by themselves define the quotient-aware estimand used
here.
The contribution is their combination for multiqubit conditional phase gates.  We
separate global equivalence-class membership from a local error generator, construct
a Ramsey graph estimator, derive exact relations between circular Boolean faces and
interaction support, identify when multi-target face tests are necessary but not
sufficient, and give a falsifiable path from estimated generator coordinates to a
local control update.  The exact membership criterion is converted into a
finite-shot statistical decision with prespecified tolerances, simultaneous
control, and \textsc{pass}/\textsc{fail}/\textsc{unresolved} outcomes, and the
target-face blind spot is upgraded to a complete count of the blind invariant
family.

The construction also connects three readerships that often use different
vocabularies.  For quantum characterization, verification, and validation (QCVV),
the key objects are an explicit estimand, explicit uncertainty accounting, model-rejection
tests, and an independent acceptance stage.  For quantum-computing applications,
the quotient specifies which phases a compiler or calibrated compensator may remove
without mistaking a dressed representative for a faulty gate.  For central-spin
registers, the same $Z$-string coefficients are linear combinations of conditional
transition phases, while their conversion into microwave or radio-frequency
controls requires a separate, well-conditioned response Jacobian.  We state these
translations explicitly so that the terms generator, gate error, and control
parameter never become synonyms.

The same structure exposes a resource boundary.  Joint spectator readout can
compress the number of compiled Ramsey basis settings to $4M$ per repetition depth
for an $M$-qubit phase map.  Yet an $(m+1)$-body conditional contrast combines
$2^m$ control words, and independent gate/reference acquisition gives a
fixed-total-shot variance proportional to $4^m/N_{\rm face}$.  Auditing every
target/complement context can add $n2^{n-1+x}$ outputs.  A linear setting count is
therefore not a proof of polynomial total-shot cost when interaction order or the
number of resolved contexts grows.  We report parameters, settings, attempted
circuits, accepted shots, reference shots, and target-gate applications separately.

The remainder of the article defines the estimation model and equivalence-class
specification in
\cref{sec:generator}, derives the Ramsey--Walsh and face estimators in
\cref{sec:learning}, and treats covariance, resources, and fidelity in
\cref{sec:statistics}.  Section~\ref{sec:control} states the additional Jacobian
conditions needed to map an estimated generator to control parameters.
Section~\ref{sec:numerics} reports the estimator-validation studies and an
end-to-end multiplexed-acquisition benchmark.  We close with the
relationship to general Hamiltonian learning and gate characterization, the
requirements for experimental deployment, and explicit limitations.

\section{Target-relative commuting error generator}
\label{sec:generator}

\subsection{Successful-block model and its validity boundary}

Let $M$ qubits span a computational space of dimension $d=2^M$, with bit strings
$z\in\{0,1\}^M$.  Projection of an implemented operation onto this space gives the
successful computational block $K$, namely the block retained by the declared
heralding or subspace-selection rule.  Here an \emph{accepted shot} is an attempted
circuit whose declared herald or subspace-resolved readout satisfies that
prespecified success rule;
its acceptance probability is reported rather than absorbed into a conditional
fidelity.  The estimation model is used only when independent off-diagonal-transfer
tests and coherence diagnostics support
\begin{equation}
K=\sum_z a_z\e^{i\phi(z)}\proj{z},\qquad 0\le a_z\le1.
\label{eq:successful-block}
\end{equation}
The phase unitary is $D_\phi=\sum_z\e^{i\phi(z)}\proj z$.  Equation
\eqref{eq:successful-block} is not imposed by data processing: appreciable
off-diagonal transfer, unresolved leakage, or gate-to-gate nonstationarity triggers
model rejection.  The amplitudes $a_z$ are retained in the leakage-aware projected
fidelity and cannot be
removed by phase equivalence.

Write the fixed diagonal target as
\begin{equation}
U_{\rm tar}=\sum_z\e^{i\theta_{\rm tar}(z)}\proj z.
\end{equation}
The target-relative phase unitary is
\begin{equation}
U_{\rm err}=U_{\rm tar}^\dagger D_\phi
=\sum_z\e^{i\epsilon(z)}\proj z.
\label{eq:uerr}
\end{equation}
Only $\epsilon$ modulo $2\pi$ is observable from one use of the gate.  We therefore
fix a reference word $z_0$, a phase-estimation window, and an unwrapping rule before
coefficient estimation.  The branch is accepted only when every inferred relative
phase lies strictly inside that interval and all declared depth-consistency tests
pass.  A boundary hit or a competing phase alias triggers \textsc{model rejection}
rather than an arbitrary logarithm-branch choice.

\subsection{Walsh coordinates and the branch-local logarithm}

For $S\subseteq[M]$ define
\begin{equation}
\chi_S(z)=(-1)^{\sum_{j\in S}z_j},\qquad
Z_S\ket z=\chi_S(z)\ket z,
\end{equation}
and the Walsh coefficient
\begin{equation}
\widehat\epsilon_S=2^{-M}\sum_z\chi_S(z)\epsilon(z).
\label{eq:walsh-transform}
\end{equation}

\begin{theorem}[Branch-local generator]
\label{thm:generator}
Inside the accepted logarithm chart,
\begin{equation}
H_{\rm err}=\frac{i}{T}\log U_{\rm err}
=\sum_{S\subseteq[M]}J_SZ_S,\qquad
J_S=-\frac{\widehat\epsilon_S}{T}.
\label{eq:herr}
\end{equation}
After fixing global phase, the nonempty-support coefficients are unique.  Adding a
constant to $\epsilon(z)$ changes only $J_\varnothing$.
\end{theorem}

The proof is Walsh orthogonality plus commutativity of all $Z_S$ and is given in
\cref{app:proofs}.  The adjective \emph{stroboscopic} is essential: $H_{\rm err}$ is
a logarithm of one target-relative gate cycle.  It need not equal the time-dependent
Hamiltonian that generated the pulse, and different microscopic errors can induce
the same $H_{\rm err}$.

\subsection{Target specifications and quotient coordinates}

Let $\cA_{\rm allow}$ be a specified collection of Pauli-$Z$ supports and define
\begin{align}
\mathfrak g_{\rm allow}&=\operatorname{span}_{\mathbb R}
\{Z_S:S\in\cA_{\rm allow}\},\\
\cG_{\rm allow}&=\exp(-iT\mathfrak g_{\rm allow}).
\label{eq:allowed-group}
\end{align}
The target contract is the ordered pair
\begin{equation}
\mathcal T=(U_{\rm tar},\cG_{\rm allow}),\qquad
[U_{\rm tar}]_{\cG}=\{VU_{\rm tar}:V\in\cG_{\rm allow}\}.
\label{eq:target-contract}
\end{equation}
Orthogonal projection in the Walsh basis gives
\begin{equation}
H_{\rm allow}=\Pi_{\rm allow}H_{\rm err},\qquad
H_{\rm forbid}=(\Id-\Pi_{\rm allow})H_{\rm err}.
\label{eq:quotient-split}
\end{equation}

Let $\mathcal C$ denote the selected injective logarithm chart after fixing global
phase, and define only its chart-compatible part of the allowed group,
\begin{equation}
\cG_{\rm allow}^{\mathcal C}
=\{\exp(-iTH):H\in\mathfrak g_{\rm allow},\ -TH\in\mathcal C\}.
\label{eq:local-allowed-set}
\end{equation}
It is generally a local set rather than the full subgroup
$\cG_{\rm allow}$.  This distinction is essential because wrapping the individual
eigenphases of an allowed unitary can create forbidden coefficients in its selected
real logarithm.

\begin{theorem}[Chart-local tangent criterion]
\label{thm:quotient}
For a target-relative unitary whose selected logarithm lies in $\mathcal C$,
\begin{equation}
U_{\rm err}\in\cG_{\rm allow}^{\mathcal C}
\quad\Longleftrightarrow\quad H_{\rm forbid}=0.
\label{eq:local-membership}
\end{equation}
\end{theorem}

The theorem supplies local feedback coordinates, not a global group-membership
test.  Global membership must be invariant under replacing any eigenphase by that
phase plus $2\pi k_z$.  Useful contracts include global phase only, global plus
single-qubit virtual-$Z$ frames, all lower-weight diagonal dressing, or the subgroup
of phases that a particular compiler can demonstrably compensate.  These choices
must carry different metric labels.

\subsection{Generalized shared-control \texorpdfstring{$C^mZ^{\otimes n}$}{CmZxn}}
\label{sec:generalized-family}

Let $C=\{c_1,\ldots,c_m\}$ be common controls, let
$\mathcal T_{\rm tgt}=\{t_1,\ldots,t_n\}$ be targets, and define
\begin{equation}
n_j=\frac{\Id-Z_j}{2},\qquad P_C=\prod_{c\in C}n_c.
\end{equation}
The standard target unitary is
\begin{equation}
U_{m,n}^{\rm std}=\prod_{j=1}^n\exp(i\pi P_Cn_{t_j}),
\label{eq:standard-cmzn}
\end{equation}
with unwrapped phase polynomial
\begin{equation}
\theta_{m,n}^{\rm std}(c,t)=
\pi\!\left(\prod_{i=1}^{m}c_i\right)\sum_{j=1}^{n}t_j.
\label{eq:standard-polynomial}
\end{equation}
The generalized target specification used below declares all diagonal Pauli-$Z$
terms of weight below $m+1$ to be equivalent:
\begin{align}
\mathfrak g_{\le m}&=\operatorname{span}_{\mathbb R}
\{Z_S:|S|\le m\},\\
\cG_{\le m}&=\exp(-iT\mathfrak g_{\le m}),\\
[U_{m,n}^{\rm std}]_{\le m}&=
\{VU_{m,n}^{\rm std}:V\in\cG_{\le m}\}.
\label{eq:generalized-class}
\end{align}
Because $n_j=(\Id-Z_j)/2$, Boolean polynomial degree and Pauli-$Z$ weight span the
same nested diagonal spaces.  To obtain a global criterion, write the relative
eigenvalue as $u(z)=\exp[i\epsilon(z)]$ and, for every
$T\subseteq[M]$, define the multiplicative M\"obius coefficient
\begin{equation}
\Mob_T[u]=\prod_{R\subseteq T}
u(\bm 1_R)^{(-1)^{|T|-|R|}},
\label{eq:mobius-invariant}
\end{equation}
where $\bm1_R$ has ones on $R$ and zeros elsewhere.  Unlike a real logarithm,
$\Mob_T$ is unchanged by every state-dependent $2\pi$ phase relabeling.

\begin{theorem}[Global low-locality membership]
\label{thm:global-membership}
For the generalized class in \cref{eq:generalized-class},
\begin{equation}
U_{\rm impl}\in[U_{m,n}^{\rm std}]_{\le m}
\quad\Longleftrightarrow\quad
\Mob_T[u]=1\quad\forall |T|\ge m+1.
\label{eq:global-membership}
\end{equation}
\end{theorem}

Equivalently, the branch-independent residuals are
$\mu_T=\arg\Mob_T[u]\in(-\pi,\pi]$.  Multiplicative M\"obius inversion reconstructs
every $u(z)$ from all $\Mob_T[u]$; hence vanishing high-degree coefficients is both
necessary and sufficient.  The proof is given in \cref{app:proofs}.  In a
chart-compatible neighborhood, \cref{thm:global-membership} and
\cref{thm:quotient} agree, but only the former tests the full generalized class.

For $CCCZ$, $m=3,n=1$.  A realized phase map is therefore a perfect
\emph{generalized phase representative} exactly when the one four-bit M\"obius
invariant of the target-relative unitary equals one.  Arbitrary zero- through
three-local diagonal phases are allowed by this contract.  For example,
\begin{equation}
V=\bigotimes_{j=1}^{4}\diag(1,\e^{4\pi i/5})
\label{eq:large-allowed-example}
\end{equation}
is allowed, so $VU_{\rm CCCZ}$ must pass.  Its principal eigenphases nevertheless
have a spurious four-body Walsh coefficient $-\pi/2$; a test based only on
$H_{\rm forbid}$ would reject it even though $\Mob_{[4]}=1$.  The gate may still
have less than unit strict fidelity to the standard target unitary, and leakage keeps the
projected generalized fidelity below one.  Thus ``$100\%$ generalized $CCCZ$'' must
never be abbreviated to ``$100\%$ $CCCZ$ gate fidelity'' without the quotient and
survival qualifiers.

\section{Ramsey--Walsh learning and completeness}
\label{sec:learning}

\subsection{Edge phases and graph reconstruction}

Choose an oriented graph $G=(V,E)$ whose vertices are computational-basis states.  A
conditional Ramsey experiment on edge $e=(u,v)$ estimates
\begin{equation}
\Delta_e=\epsilon(v)-\epsilon(u)+\xi_e.
\label{eq:edge-model}
\end{equation}
With incidence matrix $B$, $\bm\Delta=B\bm\epsilon+\bm\xi$.  If $G$ is connected,
$\rank B=d-1$ and the only null direction is a constant phase.  Deleting the fixed
reference vertex gives $B_r$ and the generalized least-squares estimator
\begin{align}
A&=(B_r^TW B_r)^{-1}B_r^TW,\\
\widehat{\bm\epsilon}_r&=A\widehat{\bm\Delta},\\
\Cov(\widehat{\bm\epsilon}_r)&=A\Cov(\widehat{\bm\Delta})A^T.
\label{eq:graph-gls}
\end{align}
Square-loop sums must vanish for a scalar phase field.  Significant loop residuals
are therefore model-validation statistics, not additional Hamiltonian coefficients.

For each qubit direction, preparing the endpoint superposition and jointly reading
all spectator outcomes bins every parallel hypercube edge in the same basis setting.
Gate $X/Y$ quadratures and equal-duration reference $X/Y$ quadratures give $4M$
compiled basis settings per repetition depth.  This compression assumes resolved
joint spectator readout; it does not remove the shot cost of low-probability bins.

\subsection{Multi-depth circular likelihood}

At repetition depth $q$, a binary Ramsey quadrature has the local form
\begin{equation}
p_+^{(q)}(\delta)=\frac{1+V_q\cos(q\delta+\psi_q)}{2},
\label{eq:ramsey-probability}
\end{equation}
with a sine quadrature obtained by a $\pi/2$ analysis shift.  Gate and reference
counts are fitted jointly within a fixed circular phase-estimation window.  We use
$q\in\{1,2,4\}$: $q=1$ anchors the branch, while the larger depths increase local
information and test the assumption of linear phase accumulation.  Repeated-gate
phase amplification is related to robust phase-estimation calibration
\cite{kimmel2015rpe,higgins2007phase}; adaptive time choice is a distinct design problem
\cite{deneeve2025timeadaptive}.

The estimator reports the likelihood minimum, conditional curvature standard error,
alias separation, boundary flag, and a depth-residual test.  A nonstationary phase
contribution present at only one repetition depth is not averaged into a more precise
estimate: it should increase the depth-consistency residual and trigger rejection of
the repeated-gate model.  Reference circuits are interleaved, and their shots are
accounted for separately so that shared drift and reference noise enter the
covariance rather than being silently discarded.

\subsection{Circular faces and local generator supersets}

Here a Boolean \emph{face} is the subcube obtained by varying a selected set of
qubits while holding all complementary spectator bits fixed; \emph{circular} means
that the resulting phase contrast is defined modulo $2\pi$.
For nonempty $A\subseteq[M]$, fix a complementary spectator word $r$ and define the
oriented face difference
\begin{equation}
D_A\epsilon(r)=\sum_{x\in\{0,1\}^{A}}
(-1)^{|A|-|x|}\epsilon(x,r).
\label{eq:face-definition}
\end{equation}

\begin{theorem}[Face--Walsh identity]
\label{thm:face-walsh}
For every real phase field in the accepted chart,
\begin{equation}
D_A\epsilon(r)=(-1)^{|A|}2^{|A|}
\sum_{S\supseteq A}\widehat\epsilon_S\chi_{S\setminus A}(r).
\label{eq:face-walsh}
\end{equation}
Consequently, all terms supported on proper subsets of $A$ cancel exactly.  A Walsh
transform over the spectator word recovers every coefficient whose support contains
$A$.
\end{theorem}

The physical face observable is instead the multiplicative contrast
\begin{align}
\Mob_A[u;r]&=\prod_{x\in\{0,1\}^{A}}
u(x,r)^{(-1)^{|A|-|x|}},\nonumber\\
\Delta_A^{\rm circ}(r)&=\arg\Mob_A[u;r].
\label{eq:circular-face}
\end{align}
It obeys $\Delta_A^{\rm circ}(r)=\wrap[D_A\epsilon(r)]$ for every choice of phase
representatives.  Thus \cref{eq:face-walsh} describes which coefficients are learned
inside an accepted real chart, whereas \cref{eq:circular-face} is the
branch-independent gate observable.

For target $t_j$ in \cref{eq:standard-cmzn}, take
$A_j=C\cup\{t_j\}$ and define
\begin{align}
\Phi_j(r)&=\arg\Mob_{A_j}[u_{\rm impl};r],\nonumber\\
\delta_j(r)&=\wrap[\Phi_j(r)-\pi].
\label{eq:target-delta}
\end{align}
Every allowed term of weight at most $m$ cancels, so $\delta_j(r)$ is the signed
deviation of the intended factor conditional phase.  Spectator dependence resolves
strict supersets of $A_j$.  This is the precise version of comparing the phase on
$\ket{11\cdots1}$ with ``the sum of the other phases'': the sum must be the oriented
Boolean finite difference, with alternating signs, not an unsigned average.

\subsection{Why correct factors need not imply a correct generalized gate}

Equations~\eqref{eq:face-walsh} and \eqref{eq:circular-face} show that all
designated target faces audit
\begin{equation}
\bigcup_{j=1}^{n}\{S:S\supseteq A_j\}.
\label{eq:face-visible-set}
\end{equation}
The complete global audit in \cref{eq:global-membership} instead contains every
$T$ with $|T|\ge m+1$.  These sets coincide for one target with no external
spectator, because $A_1$ is the only support of weight $m+1$ or higher.

\begin{proposition}[Multi-target blind support]
\label{prop:blind-support}
Let $n\ge2$ and choose any $c_*\in C$.  The support
\begin{equation}
S_*=(C\setminus\{c_*\})\cup\{t_1,t_2\}
\label{eq:blind-support}
\end{equation}
has weight $m+1$ and is forbidden under $\cG_{\le m}$, but contains no complete
$A_j$.  For an error $\epsilon(z)=a\chi_{S_*}(z)$, every designated
$\delta_j(r)$ vanishes, while the global invariant on $S_*$ is nontrivial whenever
$2^{m+1}a\notin2\pi\mathbb Z$.  In particular, every sufficiently small nonzero
$a$ violates generalized-gate membership while remaining invisible to the target
faces.
\end{proposition}

The same construction with one target and one external spectator creates a blind
support even when $n=1$.  Hence a reduced target-face calibration measurement and
the complete forbidden-spectrum audit establish different statements:
\begin{align}
\delta_j(r)=0\ \forall j,r
&\ \Longrightarrow\
\substack{\text{designated factors}\\\text{are correct}},\\
\Mob_T[u]=1\ \forall |T|\ge m+1
&\ \Longleftrightarrow\
\substack{\text{complete generalized}\\\text{phase membership}}.
\label{eq:two-claims}
\end{align}

\begin{corollary}[Complete blind family]
\label{cor:blind-family}
For a shared-control register with $m\ge1$ controls, $n\ge1$ targets, and
$x\ge0$ external spectators, let
$\mathcal H_m=\{S\subseteq[M]:|S|\ge m+1\}$ and partition it as
\begin{equation}
\mathcal V=\bigcup_{j=1}^{n}\{S\in\mathcal H_m:A_j\subseteq S\},\qquad
\mathcal B=\mathcal H_m\setminus\mathcal V.
\label{eq:blind-family}
\end{equation}
Then $\mathcal V$ consists exactly of the supports that contain all $m$ controls
and at least one target, so
\begin{equation}
|\mathcal V|=2^x(2^n-1),\qquad
|\mathcal B|=\sum_{\ell=m+1}^{M}\binom{M}{\ell}-2^x(2^n-1).
\label{eq:blind-count}
\end{equation}
In particular $\mathcal B=\varnothing$ if and only if $n=1$ and $x=0$.  Every
$S\in\mathcal B$ is an independent forbidden invariant coordinate that is
invisible to every designated target face.
\end{corollary}

A full phase map evaluates $\mathcal B$ with $|\mathcal B|$ classical
M\"obius operations and no new compiled settings.  Covering $\mathcal B$ from
target-face data alone requires additional observation designs, because each
blind support either omits at least one control or contains no target.  The
counting is exact for
independent invariant coordinates; it claims no circuit-minimal acquisition.
For $C^2Z^{\otimes2}$ it gives $L_{\rm global}=5$, $|\mathcal V|=3$, and
$|\mathcal B|=2$---the two supports $(C\setminus\{c_*\})\cup\{t_1,t_2\}$ for
$c_*\in C$---so one supplementary support check exhibits only one blind
coordinate.

\subsection{Model validation and decision logic}

No generator estimate is reported or passed to feedback solely because a phase fit
converged.  The prespecified model-validation tests check computational transfer,
aggregate and state-resolved leakage, successful-amplitude nonuniformity, Ramsey
visibility, graph loop closure, logarithm-branch capture, and linear depth scaling.
Readout-confusion correction is
accepted only when a separately calibrated confusion model improves prespecified
diagnostic metrics, and gate/reference pairing is preserved under shared drift.

The decision logic is organized in three layers.  On the model-validity layer,
the named validation tests return \textsc{model checks passed} or
\textsc{model rejected}; a rejected model does not proceed to phase-level
decisions.  On the statistical-decision layer, each prespecified claim is one
of
\begin{enumerate}
\item \textsc{pass}: every simultaneous tolerance interval for the claim lies
entirely inside the declared tolerance;
\item \textsc{fail}: at least one coordinate's simultaneous interval lies
entirely outside the declared tolerance;
\item \textsc{statistically unresolved}: an interval intersects the tolerance
boundary, so the data neither certify nor exclude the claim.
\end{enumerate}
The claim labels are \emph{targeted phase}, \emph{generalized quotient},
\emph{strict (compensated)}, and \emph{projected leakage-aware metric}; a
strict claim additionally requires absence or independently validated
compensation of the allowed dressing.  A targeted-phase \textsc{pass} never
promotes to a generalized-quotient \textsc{pass} for a multi-target or
externally spectated register, and the absence of a detected defect is not
upgraded to a \textsc{pass} without a tolerance test.  \textsc{unresolved} is
a statistical outcome, not a synonym for \textsc{model reject}; the rules are
stated in \cref{sec:finite-shot-decisions}.

\section{Statistics, complexity, noise, and fidelity}
\label{sec:statistics}

This section answers four operationally different questions: which phase functional
is identifiable, how its uncertainty propagates, how many experimental resources the
stated design consumes, and which gate-quality statement follows.  Keeping those
questions separate is especially important in QCVV: a short setting list is not a
small shot budget, and a small conditional-phase error is not by itself a
leakage-aware gate fidelity.

\subsection{Linear covariance propagation}

Once the logarithm-branch selection and the graph-consistency model have both passed
their validation tests, every Walsh coefficient and local face contrast is a linear
functional of the reconstructed phase vector.  Let $E_r$
insert the fixed zero reference coordinate into $\widehat{\bm\epsilon}_r$, let $W_M$
be the normalized Walsh matrix, let $P_F$ select forbidden rows, and let $D_A$
denote the signed face row.  From \cref{eq:graph-gls},
\begin{align}
\widehat{\bm h}&=W_ME_rA\widehat{\bm\Delta},\\
\Cov(\widehat{\bm h})&=W_ME_rA C_\Delta A^TE_r^TW_M^T,\\
\Cov(\widehat{\bm h}_F)&=P_F\Cov(\widehat{\bm h})P_F^T,\\
\Var(\widehat D_A)&=D_AE_rA C_\Delta A^TE_r^TD_A^T.
\label{eq:linear-covariance}
\end{align}
Here $\bm h$ denotes the phase coefficients $\widehat\epsilon_S$; division by
$-T$ converts their covariance to that of $J_S$.  Shared references, fitted
visibilities, readout correction, and time-block correlations must enter
$C_\Delta$.  Treating all reconstructed vertices as independent after graph
integration is generally incorrect.

A scalar functional $\ell^T\bm h$ is identifiable from a restricted design matrix
$X$ if and only if $\ell$ lies in the row space of $X$, equivalently
\begin{equation}
\rank X=\rank\!\begin{pmatrix}X\\ \ell^T\end{pmatrix}.
\label{eq:row-space-test}
\end{equation}
This test is used before concluding that a reduced calibration measurement identifies a
particular face or generator coordinate.

\subsection{Finite-shot global quotient decisions}
\label{sec:finite-shot-decisions}

\Cref{thm:global-membership} is an exact statement about a known phase map.  A
finite-shot decision must instead bind every forbidden coordinate to a
prespecified physical tolerance under simultaneous statistical control.  Let
the high-order audit family be
\begin{align}
\mathcal H_m&=\{T\subseteq[M]:|T|\ge m+1\},\nonumber\\
L_{\rm global}&=|\mathcal H_m|=\sum_{\ell=m+1}^{M}\binom{M}{\ell}.
\label{eq:global-family}
\end{align}
Inside the selected local circular chart, the residual coordinates
$\widehat\mu_T=\arg\Mob_T[\widehat u]$ assemble as
\begin{equation}
\widehat{\boldsymbol\mu}=D_{\mathcal H}\widehat{\bm\epsilon},
\qquad
C_\mu=D_{\mathcal H}C_\epsilon D_{\mathcal H}^{T},
\label{eq:mu-covariance}
\end{equation}
where $\widehat{\bm\epsilon}$ and $C_\epsilon$ come from
\cref{eq:graph-gls,eq:linear-covariance}; the $\arg\Mob_T$ observables are
insensitive to state-dependent $2\pi$ relabelings.
\Cref{eq:mu-covariance} is a local-chart approximation, conditioned on branch
capture and the depth-consistency checks.

The statistical acceptance contract augments the physical target contract
$\mathcal T$ of \cref{eq:target-contract} with a prespecified tolerance vector
$\bm\tau$---one tolerance per forbidden coordinate---a familywise level
$\alpha$, and the three-outcome simultaneous rule below.  These statistical
quantities are fixed before release data are analyzed; $\bm\tau$ does not alter
the physical equivalence class $[U_{\rm tar}]_{\mathcal G}$.  A
conservative, correlation-agnostic simultaneous critical value is
\begin{equation}
c_\alpha=\Phi^{-1}\!\left(1-\frac{\alpha}{2L_{\rm global}}\right),
\label{eq:simultaneous-critical}
\end{equation}
with $\Phi$ the standard normal cumulative distribution.  Writing
$\sigma_T=\sqrt{[C_\mu]_{TT}}$, the finite-shot decision under a valid model is
three-valued:
\begin{itemize}
\item \textsc{generalized pass}: for every $T\in\mathcal H_m$,
$|\widehat\mu_T|+c_\alpha\sigma_T\le\tau_T$;
\item \textsc{generalized fail}: for at least one $T$,
$|\widehat\mu_T|-c_\alpha\sigma_T>\tau_T$;
\item \textsc{statistically unresolved}: otherwise.
\end{itemize}
Correlation is carried by $C_\mu$, and the union bound keeps the familywise
error of the \textsc{pass}/\textsc{fail} claims at most $\alpha$ in the
local-Gaussian regime for any dependence structure; joint Gaussian
max-statistics or parametric bootstrap critical values may raise power, but
only if prespecified.  \textsc{model reject} is decided on a different layer
and preempts all three outcomes.

A fixed per-coordinate $z$-score threshold of the kind used in
\cref{sec:numerics} is a defect-detection screening rule, not an acceptance
test: failing to detect a defect does not by itself license a
\textsc{pass}.

\subsection{Precision of an arbitrary phase invariant}

Let $q_z\in\mathbb Z$ define a multiplicative phase invariant and consider its
local circular coordinate
$\vartheta_q=\wrap[\sum_zq_z(\phi_z^{(g)}-\phi_z^{(r)})]$ away from the branch cut.
If the state/arm blocks are independent and have information values $I_z^{(a)}$, the
ordinary scalar Cram\'er--Rao inequality gives
\begin{equation}
\Var(\widehat\vartheta_q)\ge
\sum_zq_z^2\left[\frac{1}{I_z^{(g)}}+\frac{1}{I_z^{(r)}}\right].
\label{eq:general-contrast-crb}
\end{equation}
For $I_z^{(a)}=N_z^{(a)}\Gamma_z^{(a)}$ and a fixed total number $N$ of
attempts, the continuous optimal allocation and bound are
\begin{align}
N_z^{(a)}&\propto\frac{|q_z|}{\sqrt{\Gamma_z^{(a)}}},\\
\Var_{\min}(\widehat\vartheta_q)&=
\frac{1}{N}\left[
\sum_{z,a}\frac{|q_z|}{\sqrt{\Gamma_z^{(a)}}}
\right]^2.
\label{eq:general-optimal-allocation}
\end{align}
The result applies to a selected local coordinate of a circular invariant.  It does
not replace branch control or a multiple-testing correction.

\subsection{Fixed-design Fisher bound for one target face}

We first count \emph{attempted circuit executions}.  Let $N_{cq}^{(a)}$ be the
attempts assigned to control word $c$, depth $q$, and arm $a\in\{g,r\}$ for gate
and reference.  Let $s_{cq}^{(a)}$ be the probability that an attempt enters the
heralded successful block and $V_{cq}^{(a)}$ its conditional Ramsey visibility.
We assume here that the heralding probability itself carries no information about
the phase estimand; survival remains a separate diagnostic.

For \cref{eq:ramsey-probability}, one attempted shot has Fisher information
\begin{equation}
\mathcal I_{cq}^{(a)}(\delta)=
s_{cq}^{(a)}
\frac{q^2[V_{cq}^{(a)}]^2\sin^2\vartheta_{cq}^{(a)}}
{1-[V_{cq}^{(a)}]^2\cos^2\vartheta_{cq}^{(a)}},
\quad
\vartheta_{cq}^{(a)}=q\delta+\psi_{cq}^{(a)}.
\label{eq:fisher}
\end{equation}
Its maximum-slope value is $s_{cq}^{(a)}q^2[V_{cq}^{(a)}]^2$.  A coarse
$X/Y$ pair can locate the phase before a local analysis axis is placed near this
maximum-slope point; retaining both quadratures instead simply uses the exact information
in \cref{eq:fisher}.

\begin{theorem}[Fixed-design target-face precision]
\label{thm:face-crb}
Suppose the successful-block, logarithm-branch, and linear-depth models pass, and the
gate/reference/bin blocks are independent.  At maximum slope define
\begin{equation}
I_c^{(a)}=\sum_{q}N_{cq}^{(a)}s_{cq}^{(a)}q^2
[V_{cq}^{(a)}]^2.
\label{eq:bin-information}
\end{equation}
Every locally unbiased regular estimator of the signed face contrast satisfies
\begin{equation}
\Var(\widehat\delta_A)\ge
\sum_{c\in\{0,1\}^{m}}
\left[\frac{1}{I_c^{(g)}}+\frac{1}{I_c^{(r)}}\right].
\label{eq:gate-ref-variance}
\end{equation}
\end{theorem}

The proof is the scalar Cram\'er--Rao inequality for a block-diagonal Fisher
matrix; it is given in \cref{app:complexity-proofs}.  If survival, visibility, or a
confusion matrix is estimated from the same data, the corresponding nuisance
parameters enter through the Fisher-matrix Schur complement and cannot improve
\cref{eq:gate-ref-variance}.

For a single depth with common $s_q,V_q$, distribute a total of $N_{\rm face}$
attempts uniformly over the $2^m$ words and equally between the two arms.
\Cref{thm:face-crb} gives
\begin{equation}
\boxed{\;
\Var(\widehat\delta_A)\ge
\frac{4^{m+1}}{N_{\rm face}s_qq^2V_q^2}\; .}
\label{eq:rare-bin-law}
\end{equation}
Thus this state-resolved acquisition needs
$N_{\rm face}=\Theta[4^m/(s_qq^2V_q^2\epsilon^2)]$ for local standard deviation
$\epsilon$ when a regular efficient estimator reaches the inverse-Fisher scaling.
The coefficient includes both gate and reference attempts.  This is a
protocol-level local-asymptotic law, not a minimax lower bound over adaptive,
entangled, collective, or differently multiplexed experiments.

If the arms have per-attempt informations $\Gamma_g$ and $\Gamma_r$, minimizing
$4^m[(N_g\Gamma_g)^{-1}+(N_r\Gamma_r)^{-1}]$ at fixed
$N_g+N_r=N_{\rm face}$ gives
\begin{align}
\frac{N_g}{N_r}&=\sqrt{\frac{\Gamma_r}{\Gamma_g}},\nonumber\\
\Var_{\min}(\widehat\delta_A)&=
\frac{4^m}{N_{\rm face}}
\left(\Gamma_g^{-1/2}+\Gamma_r^{-1/2}\right)^2.
\label{eq:optimal-arm-allocation}
\end{align}
The less informative arm receives more attempts.

\subsection{From one face to all targets}

One fixed target face has no intrinsic $n$ penalty once its spectator context is
prepared.  An exhaustive state-resolved audit is different.  For each of $n$
targets it resolves the other $n-1$ targets and $x$ external spectators, giving
\begin{equation}
L=n\,2^{n-1+x}
\label{eq:face-family-size}
\end{equation}
target--context faces.  If each face receives a separate budget, the protocol-level
cost is
\begin{equation}
N_{\rm total}=
\Theta\!\left[
\frac{n2^{n-1+x}4^m}
{\Gamma\,\epsilon^2}
\right],
\qquad \Gamma=s_qq^2V_q^2 .
\label{eq:mn-shot-scaling}
\end{equation}
Shared reference data or structured joint fits change the covariance and may reuse
shots, so \cref{eq:mn-shot-scaling} is an acquisition count, not a universal
necessity theorem.

For a two-sided familywise false-positive budget $\alpha$, miss probability
$\beta$, and signal magnitude $\Delta$, a Bonferroni-normal design gives the
local-asymptotic sufficient allocation
\begin{equation}
N_{\rm face}\ge
\frac{4^{m+1}}{\Gamma\Delta^2}
\left[
z_{1-\alpha/(2L)}+z_{1-\beta}
\right]^2.
\label{eq:simultaneous-detection}
\end{equation}
It makes the confidence target explicit; it is not claimed to be the optimal
simultaneous test.

\subsection{Model dimension, settings, and classical work}

Let $d=2^M$ and let $D_Q$ be the number of repetition depths.  The dimensions in
\cref{tab:complexity} explain where the structural reduction occurs.  A dense
diagonal generator is exponentially smaller than a general channel but still has
$2^M-1$ physical phase coordinates.  Fixed-$k$ locality makes the parameter count
polynomial, but the present exhaustive acquisition does not automatically inherit
that scaling.

\begin{table*}[t]
\caption{Distinct resource and complexity accounts for an $M$-qubit operation.  Coordinate counts
are real dimensions after the stated gauge or promise; they are not finite-shot
sample bounds.}
\label{tab:complexity}
\begin{ruledtabular}
\begin{tabular}{p{0.25\textwidth}p{0.25\textwidth}p{0.40\textwidth}}
Object or design & Count & Interpretation\\
\hline
General CPTP map & $d^4-d^2=\Theta(16^M)$ &
Affine process-model dimension after trace preservation\\
General unitary channel & $d^2-1=\Theta(4^M)$ &
Unitary promise with channel-global phase removed\\
Dense diagonal generator & $d-1=2^M-1$ &
Known eigenbasis; one global-phase gauge\\
$k$-local diagonal generator &
$\sum_{\ell=1}^{k}\binom{M}{\ell}=O(M^k)$ for fixed $k$ &
Ansatz dimension only; support identifiability is still required\\
Complete global quotient audit &
$\sum_{\ell=m+1}^{M}\binom{M}{\ell}$ &
Multiplicative M\"obius invariants for complete $\cG_{\le m}$ membership\\
Multiplexed Ramsey settings & $4MD_Q$ full map; $4nD_Q$ target directions &
Joint spectator readout; rare-bin shot costs remain\\
Statewise comparator settings & $4D_Q2^M$ &
Separately compiled phase preparations used for an equal-shot comparator\\
\end{tabular}
\end{ruledtabular}
\end{table*}

Once all $2^M$ phases are available, fast Walsh and Boolean M\"obius transforms
cost $O(M2^M)$ arithmetic and $O(2^M)$ memory.  A full hypercube has
$M2^{M-1}$ edges; graph integration can exploit its sparse Laplacian, but we do not
claim an implementation-level asymptotic speedup for the present solver.  The exact
membership test in \cref{thm:global-membership} is distinct from optimizing the
global quotient fidelity away from membership, which can remain nonconvex.  General unitary
and process tomography have different measurement promises and dimensions
\cite{poyatos1997process,gutoski2014unitary}; the comparison in
\cref{tab:complexity} is therefore structural rather than a head-to-head precision
benchmark.

The finite-shot decision of \cref{sec:finite-shot-decisions} uses $L_{\rm global}$
only to size the simultaneous critical value
\cref{eq:simultaneous-critical}.  It is deliberately not converted into a
universal shot bound, because a full phase map makes many invariants share data
through a correlated covariance: a prespecified acquisition design and the
joint covariance $C_\mu$ are the correct inputs to any finite-shot power
statement.

\subsection{Noise penalties and optimal repetition depth}

Some noise mechanisms preserve the likelihood and simply reduce information.  For
per-gate Markovian dephasing $\gamma$, heralded survival loss $\ell$, symmetric
target-readout error $r_t$, and base visibility $V_0$, take
\begin{align}
V_q&=V_0(1-2r_t)\e^{-\gamma q},&
s_q&=\e^{-\ell q},\nonumber\\
\Gamma_q&=s_qq^2V_q^2
=q^2V_0^2(1-2r_t)^2\e^{-(2\gamma+\ell)q}.
\label{eq:markov-information}
\end{align}
With $a=2\gamma+\ell$, maximizing information per attempted circuit gives
$q_{\rm att}^\star=2/a$; maximizing information per target-gate application gives
$q_{\rm call}^\star=1/a$.  These continuous optima are clipped to the allowed
integer grid and by branch capture, visibility, and depth-residual tests.

An inhomogeneous quasi-static Gaussian phase offset---constant during one circuit but
redrawn independently between attempts---with standard deviation $\sigma$ contributes
visibility $\exp(-\sigma^2q^2/2)$ and hence information
$\exp(-\sigma^2q^2)$.  An offset fixed throughout a data block instead belongs in the
covariance or bias account below.  Without the linear decay, the corresponding optima are
$q_{\rm att}^\star=1/\sigma$ and
$q_{\rm call}^\star=1/(\sqrt2\,\sigma)$.  Longer coherent evolution can therefore
improve attempted-circuit cost while worsening target-gate-application cost or losing branch/model
validity; these resources cannot be interchanged silently
\cite{dutt2023active,hu2025ansatzfree}.

Readout of the context word creates a second penalty.  If $h$ context bits have
independent symmetric confusion probability $r_c$, the tensor-product confusion
matrix has
\begin{equation}
\sigma_{\min}(C_h)=(1-2r_c)^h,\qquad
\mathcal A_{\rm var}^{\rm worst}=(1-2r_c)^{-2h}.
\label{eq:readout-amplification}
\end{equation}
The variance factor assumes a known, stable matrix is inverted.  Unknown or drifting
confusion instead creates bias or model failure.  This distinction is shared with
multiqubit readout-mitigation methods based on calibrated confusion models
\cite{bravyi2021measurement}.

\subsection{Covariance geometry and non-averaging bias}

Let $\bm a$ be the $2^m$ vector of face signs and let $C_\phi$ be the covariance of
random bin phases not already included in the shot likelihood.  Linear propagation
gives
\begin{equation}
\Var_{\rm noise}(\widehat\delta_A)=\bm a^TC_\phi\bm a.
\label{eq:face-noise-covariance}
\end{equation}
Common-mode noise cancels because $\bm a^T\bm 1=0$; independent bin variance
$\sigma^2$ gives $2^m\sigma^2$; a sign-aligned rank-one covariance
$\sigma^2\bm a\bm a^T$ gives $4^m\sigma^2$.  A single scalar noise rate therefore
does not determine the face precision.

Deterministic biases are more severe.  If every bin obeys $|b_c|\le b_{\max}$,
\begin{equation}
|b_A|=\left|\sum_ca_cb_c\right|\le2^m b_{\max},
\label{eq:face-bias-bound}
\end{equation}
and the bound is tight.  A worst-case face tolerance $\epsilon$ requires
$b_{\max}\le\epsilon/2^m$; more repetitions do not remove this floor.  In the local
linear/Gaussian regime it is useful to keep the uncertainty contributions separate,
\begin{equation}
\operatorname{MSE}(\widehat\delta_A)\simeq
\Var_{\rm shot}+\bm a^TC_\phi\bm a+b_A^2 .
\label{eq:face-mse}
\end{equation}

Finally, observations $y_q=q\delta+\beta_q$ fitted to a through-origin linear law
have
\begin{equation}
\operatorname{bias}(\widehat\delta)=
\frac{\sum_qw_qq\beta_q}{\sum_qw_qq^2},\qquad
|\operatorname{bias}|\le
\frac{\sqrt{\sum_qw_q\beta_q^2}}{\sqrt{\sum_qw_qq^2}}.
\label{eq:depth-bias}
\end{equation}
The residual must still pass a separate depth-law test.  Unheralded leakage,
off-diagonal coherent transfer, nonstationarity, or a failed loop/depth validation
test does not receive a larger error bar; it triggers \textsc{model rejection}.

\subsection{Strict, quotient, and leakage-aware metrics}

For a diagonal unitary relative error, Haar averaging gives
\begin{equation}
F_{\rm strict}=
\frac{d+|\Tr U_{\rm err}|^2}{d(d+1)}.
\label{eq:strict-fidelity}
\end{equation}
The exact quotient definition is
\begin{equation}
F_{\rm gen}=\max_{V\in\cG_{\rm allow}}
F_{\rm avg}(D_\phi,VU_{\rm tar}).
\label{eq:generalized-fidelity}
\end{equation}
Likewise $F_{\rm comp}$ uses only a declared physically compensable subgroup.
Membership in the low-locality class is decided without a logarithm by
\cref{eq:global-membership}; evaluating the global maximum in
\cref{eq:generalized-fidelity} away from membership can still be a nonconvex phase
optimization.  When a local branch is captured, cancelling $H_{\rm allow}$ supplies
the reproducible candidate
\begin{align}
V_{\rm proj}&=\exp(-iT H_{\rm allow}),\\
F_{{\rm gen},\rm LB}&=F_{\rm avg}(D_\phi,V_{\rm proj}U_{\rm tar})
\le F_{\rm gen}.
\label{eq:quotient-lower-bound}
\end{align}
We do not relabel this lower bound as the global optimum without a certificate.

For the nonunitary successful block $K$ in \cref{eq:successful-block}, the projected
average overlap with a unitary target $U$ is exactly \cite{nielsen2002average}
\begin{equation}
F_{\rm proj}(K,U)=
\frac{\Tr(K^\dagger K)+|\Tr(U^\dagger K)|^2}{d(d+1)}.
\label{eq:projected-fidelity}
\end{equation}
Maximizing its phase-overlap term over $V$ leaves $\Tr(K^\dagger K)$ unchanged.
Thus neither leakage nor nonuniform successful amplitudes are quotiented away.
For uniform survival $K=\sqrt p,VU_{\rm tar}$, even perfect generalized phase
membership gives $F_{\rm proj}=p$.  Leakage diagnostics and subspace fidelity are
therefore complementary rather than interchangeable
\cite{wood2018leakage,chasseur2015leakage}.

If the factorized shared-control error model has been independently validated, with
independent phase errors $\delta_j$ only on the intended $C^mP_{t_j}$ factors and
no extra spectators, its trace is
\begin{equation}
\Tr U_{\rm err}=(2^m-1)2^n+
\prod_{j=1}^{n}(1+\e^{i\delta_j}).
\label{eq:factor-trace}
\end{equation}
Equations~\eqref{eq:strict-fidelity} and \eqref{eq:factor-trace} give an exact
closed-form fidelity expression for continuous phase errors.  It is not used when
spectator dependence or the full forbidden spectrum rules out the factorized model.

\section{From learned generator to a control update}
\label{sec:control}

This section is an illustrative interface from the structured generator to a
local control proposal: it is not an independent QCVV pass, a closed-loop
convergence theorem, or a hardware calibration result.  Learning $H_{\rm err}$
does not by itself identify a physical error source.  Let
$\bm\theta$ denote available controls and let $\bm h(\bm\theta)$ collect either all
nonempty phase coefficients (strict objective) or only forbidden coefficients
(quotient-aware objective).  Mapping these estimates to an update requires an
additional control-response model.  A local linear-response model, estimated by
small control perturbations, is
\begin{equation}
\bm h(\bm\theta+\bm u)\simeq
\bm h(\bm\theta)+J_\theta\bm u,\qquad
(J_\theta)_{S\alpha}=\frac{\partial h_S}{\partial\theta_\alpha}.
\label{eq:control-jacobian}
\end{equation}
The generator-to-control inverse problem requires a declared rank, an absolute
sensitivity floor, and a condition-number criterion.  Formally nonzero derivatives below
the floor are treated as unidentifiable.  With covariance weight $W$, ridge
$\lambda>0$, and predeclared bounds $\mathcal B$, one candidate update is
\begin{equation}
\bm u_*=\arg\min_{\bm u\in\mathcal B}
\|J_\theta\bm u+\widehat{\bm h}\|_W^2+
\lambda\|\bm u\|_2^2.
\label{eq:bounded-update}
\end{equation}
The update is only a local control proposal.  Independent data must re-evaluate the
model and the chosen target metric; an improvement in phase coefficients alone is not accepted if
the leakage-aware projected metric worsens.

\subsection{Central-spin instantiation}

Frequency-resolved central-spin registers provide a concrete diagonal-gate platform
\cite{pla2013nuclear,taminiau2014universal,bradley2019ten,abobeih2019imaging}.  Nuclear configurations
shift an electron-spin resonance, and a closed conditional electron cycle can attach
a spinor phase to selected computational-basis states.  Multitone global driving and
intersubspace dynamical decoupling offer one robust-control construction
\cite{song2026idd}; here the architecture is used only as a simulation testbed for
the learning-to-control interface.  In this language, $J_S/(2\pi)$ is an effective
conditional-frequency coordinate, a Boolean face is an alternating difference of
state-conditioned cycling phases, and $J_\theta$ maps those observable coordinates to
tone frequencies, amplitudes, or phases.  None of these effective quantities is
automatically a microscopic hyperfine or drive-Hamiltonian parameter.

The numerical model uses an electron frequency of 32\,GHz, four conditional
hyperfine shifts $\{20,40,80,160\}\,$MHz, and a 1\,\textmu s gate.  A
carrier rotating-wave conditional-block propagator supplies target-relative phases,
survival amplitudes, and leakage.  It omits measured transfer functions,
dissipation, and thermal recovery.  In the primary update, the simulated system and
the finite-difference Jacobian are generated by the same conditional-block model; a
separate sensitivity study below scales and reverses the validation simulator's
control response.  We inject common detuning, one selected-tone detuning,
active-tone gain, or selected-tone phase.  Strict fits use every nonempty Walsh coefficient;
quotient-aware fits use only $|S|\ge m+1$.  Each case receives one bounded update.

The selected-tone phase is included deliberately as a negative identifiability
control.  In this conditional-block model, it rotates a transverse drive axis but
does not materially change the estimated diagonal phase generator.  A method that
turns such a nearly null Jacobian column into a claimed phase calibration would be
overinterpreting its learned object.

\section{Numerical validation}
\label{sec:numerics}

\subsection{Simulation design and reporting}

All numerical results below are seeded simulations.  Exploratory runs were used to
identify branch-capture, low-sensitivity, and finite-replicate failure modes before
the reported parameter grids and decision rules were fixed.  The final studies
retain the replicate-level estimates needed to recompute every plotted sample
variance, RMSE, and rejection rate.  The end-to-end acquisition benchmark is
generated from count records, whereas the response-gain sweep is a deterministic
sensitivity calculation.  These numerical checks test the estimators and named
failure modes under their stated models; they are not experimental gate
certification.

\subsection{Global quotient and end-to-end acquisition}

The first check targets the branch issue directly.  We multiply an ideal
$CCCZ$ by four allowed single-qubit gates
$\diag(1,\e^{4\pi i/5})$.  The complete circular audit returns
$\max_{|T|\ge4}|\mu_T|<9.8\times10^{-16}$ rad, as required for exact generalized
membership.  Yet the captured principal logarithm contains a four-body Walsh
coefficient of magnitude $\pi/2$ and would fail the local tangent criterion.  This
is a constructive regression test against misclassifying allowed large phase
dressing as a physical generalized-gate defect.

We next simulate the measurement protocol from multinomial spectator-bin counts,
not from pre-generated phase estimates.  Each qubit direction uses gate/reference
$X/Y$ quadratures, so the multiplexed design has $4M$ compiled settings and jointly
bins all parallel hypercube edges.  At a configured budget of 524,288 attempted
circuits per phase map, integer allocation uses 524,280 attempts for $M=3,5$ and
524,288 for $M=4$; the statewise comparator uses 524,288.  Across 200 independent
replicates, the ratios of mean gauge-aligned phase-map RMSE
(multiplexed/statewise) are
$0.689$, $0.739$, and $0.794$ for $M=3,4,5$.  The corresponding setting counts are
$12/32$, $16/64$, and $20/128$.  These finite-grid results validate the implemented
joint-binning resource accounting; they do not prove a universal precision advantage.

Finally, for $C^2Z^{\otimes2}$ we inject the blind Walsh term
$0.12 Z_{c_2t_1t_2}$ rad.  Its exact global circular residual is $0.96$ rad.
This register has $m=n=2$, $x=0$, so $L_{\rm global}=5$ with $|\mathcal B|=2$;
the injected support is one of the two blind coordinates.
The screening statistic is the family-max studentized score
$\max_{T\in\mathcal H_m}|\widehat\mu_T|/\sigma_T$, evaluated from the estimated
phase map and its propagated covariance at the fixed $z=3.5$ threshold.  At 300
replicates, the complete invariant audit rejects $300/300$ (score-Wilson
$95\%$ interval $[0.987,1]$), whereas the two target-face family rejects
$1/300$, the same rate as its allowed-only target-face control.  The
allowed-only global audit rejects $2/300$ (Wilson $95\%$ interval
$[0.002,0.024]$), which is the empirical family-level false rejection of this
prespecified screening design, not a nominal $0.05$ coverage claim.

The fixed $z=3.5$ threshold is a preregistered defect-detection screening rule,
not an acceptance test: its per-coordinate two-sided level is
$4.7\times10^{-4}$, and the union bound over the five coordinates gives a
familywise level near $2.3\times10^{-3}$.  A deterministic recomputation from
the stored replicate-level records also exercises the Bonferroni-normal
simultaneous rule of \cref{eq:simultaneous-critical}
($c_{0.05}=2.576$): on the allowed-only null it rejects $14/300$
(score-Wilson $95\%$ interval $[0.028,0.077]$), consistent with the nominal
$0.05$ familywise level, and on the blind family $300/300$.  The fixed
$z=3.5$ screening rule is thus more conservative than the nominal $0.05$
familywise defect-detection rule; $300/300$
remains empirical detection at this effect size, not universal power or
certification.  Thus the simulation
checks both sides of the logical claim: the complete audit sees the defect,
while the proposed reduced target-face tests remain blind.

\begin{figure*}[t]
\centering
\includegraphics[width=\textwidth]{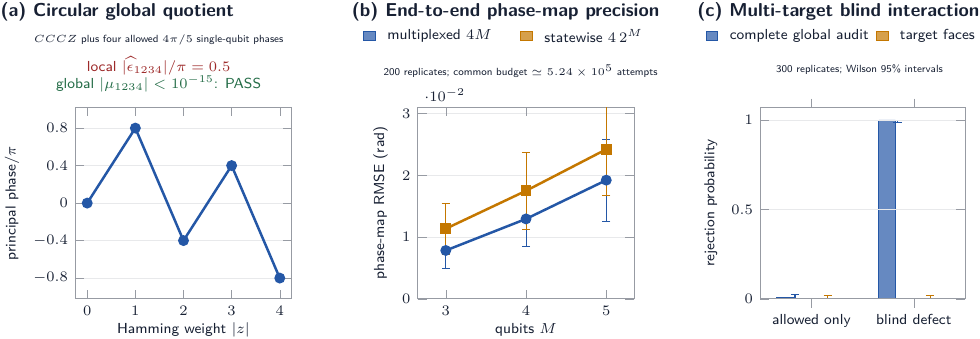}
\caption{Global-quotient and acquisition checks.  (a) Principal phases of the
allowed dressing in the $CCCZ$ alias regression: a local logarithm reports a
spurious four-body coefficient, while the branch-independent invariant passes.
(b) Gauge-aligned phase-map RMSE for end-to-end multiplexed and separately compiled
statewise acquisition under a common nominal attempted-circuit budget.  Points and
error bars are the mean and sample SD of the per-map RMSE across 200 independent
replicates; they are not standard errors.  (c) Rejection rates for the allowed-only
control and the blind-defect alternative over 300 replicates.  Vertical intervals
are score-Wilson $95\%$ binomial intervals computed from the rejection indicators.}
\label{fig:global-validation}
\end{figure*}

\subsection{Complexity and noise validation}

This validation study tests the finite-shot implementation independently of the earlier
gate-classification and central-spin studies.  For $m=1,\ldots,6$, each face receives
262,144 attempted circuit executions, shared uniformly across control words and equal gate and
reference arms, with $q=2$, $s=0.88$, and $V=0.82$.  Over 1500 fresh replicates per
point, the empirical-to-CRB variance ratios are
\[
0.956,\ 0.957,\ 1.026,\ 1.055,\ 0.962,\ 1.055,
\]
and the fitted slope of $\log_2\Var(\widehat\delta_A)$ versus $m$ is $2.022$
(analytic value $2$).  Nominal $95\%$ local-normal coverage lies between $0.947$
and $0.959$.  This validates the implemented fixed-design law over the specified grid;
it does not estimate a universal learning exponent.

\begin{figure*}[t]
\centering
\includegraphics[width=\textwidth]{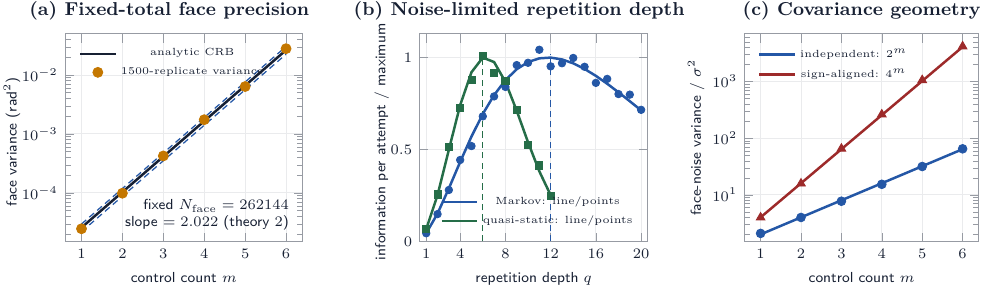}
\caption{Complexity and noise validation.  (a) Exact
\cref{eq:rare-bin-law} (solid line) and sample variances from 1500 saved
replicate-level estimates (points).  Dashed lines are the prespecified
$0.85$--$1.15$ ratio criterion, not uncertainty bars.
(b) Analytic information per attempted circuit and empirical inverse variance,
each normalized by the analytic maximum.  Blue and green vertical lines mark the
predicted integer optima $q=12$ (Markov) and $q=6$ (quasi-static); empirical optima
are $11$ and $6$.  Lines are analytic and points are empirical.
(c) Face-noise variance normalized by $\sigma^2$ for independent and sign-aligned
Gaussian covariance; points use 3000 replicates and lines are $2^m$ and $4^m$.
Common-mode noise cancels exactly and is omitted from the logarithmic axis.  No
per-point error bar is drawn because none was prespecified; all points, child
seeds, and acceptance decisions are retained in the underlying analysis record.}
\label{fig:complexity-noise}
\end{figure*}

The Markov cell uses $\gamma=0.07$ and $\ell=0.03$, so
\cref{eq:markov-information} predicts continuous optima $q=11.765$ per attempted
circuit and $q=5.882$ per target-gate application, and an attempted-circuit grid
optimum $q=12$; the empirical
grid optimum is $11$.  For the independently redrawn quasi-static dephasing model with
$\sigma=0.16$, the predicted and empirical
attempted-circuit optima are both $q=6$.  All 32 depth-cell variance ratios lie between
$0.922$ and $1.100$.

Two further checks expose effects that a depth curve alone can miss.  An explicit
independent confusion tensor with $r_c=0.08$ reaches a worst-direction variance
amplification of $16.275$ at eight context bits and matches
\cref{eq:readout-amplification} to $1.4\times10^{-16}$ absolute residual.  The
correlated-noise study gives the $2^m$ and $4^m$ laws in
\cref{fig:complexity-noise}(c), while common mode cancels.  Finally, adversarial
$1$-mrad bin biases saturate the $2^m$ bound, and an injected nonlinear depth defect
produces a $0.0248$-rad weighted residual against a $0.01$-rad threshold, triggering
\textsc{model rejection}.  These negative controls prevent more shots from being
presented as a remedy for systematic error.

\subsection{Generic observation-model validation}

The generic observation-model validation contains four complementary studies.  First, $1000$ rare-bin
replicates for $m=1,\ldots,5$ compare empirical variance with the conditional
Fisher prediction.  At $8192$ total gate shots and uniform occupancy, the five
empirical-to-conditional variance ratios are
$0.971,1.015,0.988,1.040,0.998$.  Prespecified low-count/skewed cells also retain
their intended failure rates: the point is to validate both the local law and its
count gate, not to hide invalid bins.

\begin{figure*}[t]
\centering
\includegraphics[width=\textwidth]{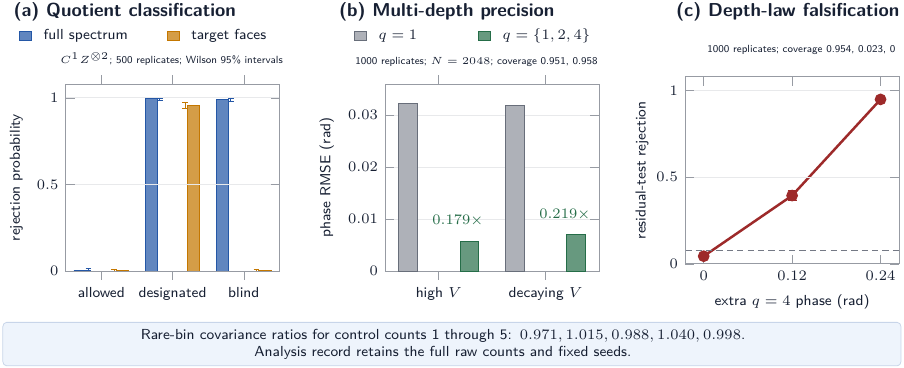}
\caption{Count-level observation-model validation.  (a) Full-spectrum versus
target-face rejection in the representative $C^1Z^{\otimes2}$ case.  Bars are
empirical rates and vertical lines are score-Wilson $95\%$ binomial intervals from
500 independent seeded replicates; each interval is computed from its corresponding
rejection indicators.
(b) Phase RMSE for $q=1$ and $q=\{1,2,4\}$ over 1000 replicates.  These bars are
RMSE summaries and do not carry invented uncertainty bars.  (c) Rejection of the
linear-depth model under an extra $q=4$ phase; intervals are score-Wilson $95\%$
intervals from 1000 Bernoulli rejection indicators.  Exact counts, seeds, and
reference-shot accounts are retained in the underlying analysis record.}
\label{fig:generic-validation}
\end{figure*}

Second, full-spectrum and designated-face tests are compared on allowed-only,
designated-forbidden, and blind-forbidden phase fields.  Each family/case uses 500
replicates, 4096 shots per quadrature, and a fixed $z=3.5$ screening threshold.
Across all four register cases every branch fit is valid.  Allowed-only false
rejection is at most $0.6\%$; designated alternatives have full-spectrum power at
least $99.6\%$ and face power at least $95.8\%$; blind alternatives have
full-spectrum power at least $99.2\%$ while target-face rejection is at most
$0.2\%$.  Panel (a) of \cref{fig:generic-validation} shows the least separated
$C^1Z^{\otimes2}$ case.

Third, for a true phase $0.035$ rad at 2048 shots per quadrature, the
$q=\{1,2,4\}$ estimator reduces RMSE from $0.03230$ to $0.005782$ rad at constant
visibility and from $0.03178$ to $0.006965$ rad under depth-decaying visibility.
Empirical-SD/reported-SE ratios are $0.965$ and $0.971$, and nominal $95\%$
coverage is $0.951$ and $0.958$.  An extra nonrepeating phase injected only at
$q=4$ demonstrates the boundary: residual-test rejection is $0.044$ under the
null, $0.394$ at $0.12$ rad, and $0.949$ at $0.24$ rad; the misspecified estimator's
coverage falls from $0.954$ to $0.023$ and $0$.

Finally, 1000-replicate model-validation tests cover transfer, aggregate
leakage, state-dependent survival, visibility loss, loop mismatch, shared drift,
and readout confusion.  Named null rejection rates lie between $0.046$ and $0.057$.
At the specified alternatives, powers are $0.832$ for $0.001$ transfer excess, $0.971$
for $0.002$ leakage excess, $0.993$ for $0.005$ state-survival excess, and $0.880$
for $0.01$ visibility loss.  Loop-mismatch power is $0.498$ at $0.1$ rad and $0.981$
at $0.2$ rad.  Pairing gate and reference counts under shared drift retains $0.947$
coverage, whereas an intentionally unpaired analysis gives $0.332$; declared readout
correction reduces both bias and RMSE at the specified confusion point.  These are
prespecified alternatives, not a universal off-model detection theorem.

The complete $M=5$ phase-map comparator uses 128 compiled quadrature settings,
524,288 circuit shots (262,144 reference shots), and 262,144 target-gate applications per
replicate.  The multi-depth scalar study uses 12 compiled quadrature settings,
24,576 circuit shots, 12,288 reference shots, and 28,672 target-gate applications.
Reporting these resources separately prevents a small setting count from being
misinterpreted as a small coherent-evolution cost.

\subsection{Learning-to-control simulation}

Each central-spin case uses 30 independently seeded pre-update Ramsey--Walsh data
sets at $q=\{1,2,4\}$ and 2048 shots per quadrature.  Strict and quotient-aware
objectives share the same pre-update counts within a replicate.  Each proposed
update is evaluated using noiseless ground truth from the same simulator model and a separately
seeded post-update phase measurement.  Every pre- and post-update phase fit is valid,
no bound is hit, and coarse/fine strict-fidelity changes remain below $1.63\times
10^{-5}$.

\begin{table*}[t]
\caption{One-step central-spin results.  Frequency updates are in Hz; gain is
fractional.  Parentheses give the sample SD across 30 noisy training replicates.
$\Delta F_{\rm proj}$ is the mean noiseless same-model projected-fidelity change and
$\Delta\|h_F\|_2$ is the mean forbidden phase-coefficient norm change.}
\label{tab:control-results}
\begin{ruledtabular}
\begin{tabular}{llrrrr}
Case & Objective & Mean update (SD) & $\|J_\theta\|_2$ &
$\Delta F_{\rm proj}$ & $\Delta\|h_F\|_2$ (rad)\\
\hline
Common detuning & strict & $+43920\ (2379)$ & $7.574\times10^{-3}$ &
$+1.0748\times10^{-3}$ & $+7.545\times10^{-4}$\\
Common detuning & quotient & $+15655\ (6599)$ & $1.950\times10^{-3}$ &
$+6.0140\times10^{-4}$ & $-2.7026\times10^{-3}$\\
Selected detuning & strict & $-55190\ (2693)$ & $7.574\times10^{-3}$ &
$+1.6719\times10^{-3}$ & $-1.0794\times10^{-2}$\\
Selected detuning & quotient & $-64626\ (8381)$ & $1.951\times10^{-3}$ &
$+1.5932\times10^{-3}$ & $-1.2641\times10^{-2}$\\
Active gain & strict & $-0.01770\ (0.00188)$ & $1.906\times10^{-4}$ &
$+4.8344\times10^{-4}$ & $-2.1141\times10^{-4}$\\
\end{tabular}
\end{ruledtabular}
\end{table*}

The common-detuning rows illustrate why the quotient changes the optimization
problem.  The strict update improves strict projected fidelity but increases the
forbidden norm, while the quotient-aware update suppresses the forbidden norm by
$2.70\times10^{-3}$ rad.  For selected detuning both objectives help both metrics,
with the quotient fit giving the larger forbidden-norm reduction.  The gain update
improves projected fidelity largely by reducing simulated leakage.

The selected-tone-phase negative controls have Jacobian sensitivities
$5.51\times10^{-8}$ (strict) and $3.79\times10^{-8}$ (quotient), both below the
fixed $10^{-6}$ floor.  Mean absolute updates are only $1.67\times10^{-5}$ and
$1.32\times10^{-6}$ rad, and metric changes are below $10^{-12}$.  They correctly
remain negative rather than being interpreted as physical-parameter recovery.

\begin{figure*}[t]
\centering
\includegraphics[width=\textwidth]{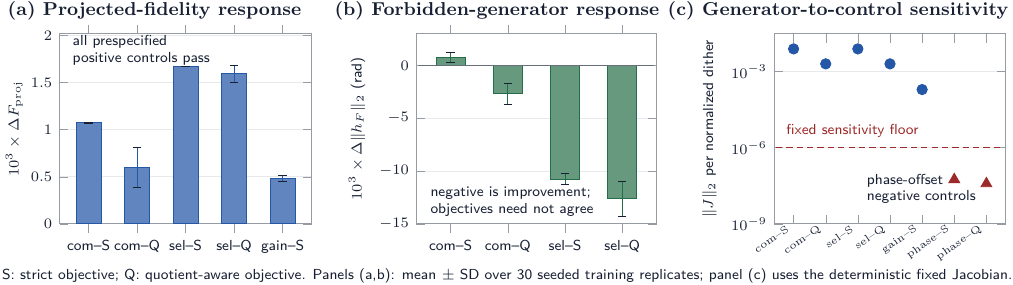}
\caption{Central-spin learning-to-control simulation.  S and Q denote strict
and quotient-aware objectives.  (a,b) Bars are mean changes and error bars are the
sample standard deviation---not the standard error---over the same 30 prespecified
training replicates; the deterministic pre-update simulator value is common to those
replicates.  The visibly different error-bar sizes follow directly from the stored
replicate outcomes.  (c) Deterministic finite-difference Jacobian sensitivities; triangles
are selected-tone-phase negative controls below the prespecified sensitivity floor.
The simulated system and Jacobian use the same conditional-block model, so this is not a
model-robust or hardware calibration result.}
\label{fig:control}
\end{figure*}

One four-spin phase-map measurement in this study comprises 192 compiled quadrature
settings, 393,216 circuit shots including 196,608 reference shots, and 458,752
target-gate applications.  A pre-update map is shared between the two objectives;
each objective has a separately seeded validation map.  The independent counts check
the measurement path, while the primary improvement estimand uses known simulator
truth.  No PTCB-like held-out benchmark was simulated in this study.

A separate deterministic sensitivity study introduces a controlled model mismatch.
The nominal $C^3Z$ common-detuning Jacobian proposes a $15.899$-kHz update, while the
validation simulator applies a response scaled by a gain
$g$.  The relative forbidden-norm reduction is $79.3\%$ at $g=1$, $39.6\%$ at
$g=0.5$, $19.8\%$ at $g=0.25$, and zero at $g=0$.  Reversing the response worsens
the norm by $39.6\%$ at $g=-0.5$ and $79.1\%$ at $g=-1$.  This isolates a concrete
calibration-map failure boundary: smaller positive gain erodes benefit, zero gain
nulls it, and a sign error reverses it.  It is not a robustness theorem for arbitrary
model mismatch, drift, or dissipation.

\section{Discussion}
\label{sec:discussion}

\subsection{What kind of Hamiltonian learning is this?}

The framework is Hamiltonian learning in a precise, restricted sense.  Its global
object is the target-relative $U(1)$ phase field of one gate cycle; only after a
logarithm branch is accepted do we interpret its real logarithm as a commuting
Pauli-$Z$ Hamiltonian.  Ramsey data provide direct access to eigenphase differences in the known
basis, so graph integration replaces the state-preparation and observable designs
needed for an unknown-basis many-body Hamiltonian.  Multiplicative M\"obius
coordinates classify the gate globally, whereas Walsh coordinates describe the
local tangent generator used for feedback.  This makes the inference transparent,
but it also means that sample-complexity results for
local Gibbs-state or real-time Hamiltonian learning
\cite{bairey2019local,haah2024learning,gu2024practical} do not transfer automatically.
In particular, the desired interaction itself can have order $m+1$, and
\cref{eq:rare-bin-law} exposes an exponential-in-$m$ conditional contrast cost.

The word \emph{query} is consequently too ambiguous for the present protocol unless
its resource is named.  One compiled setting may return many spectator-resolved
outcomes, one attempted circuit may fail the successful-block herald, and a depth-$q$
circuit applies the target gate $q$ times.  Under uniform, separate-bin acquisition,
\cref{eq:rare-bin-law} is a local Cram\'er--Rao bound and an attainable asymptotic
scaling for one face: fixed precision costs $\Theta(4^m/\Gamma\epsilon^2)$ attempted
circuits.  Under separate budgets for every target--context face,
\cref{eq:mn-shot-scaling} adds $n2^{n-1+x}$.  The latter factor is protocol accounting,
not a lower bound: joint outcome reuse, sparsity, active designs, or a different
observable can change it.  Conversely, the polynomial coordinate count of a
fixed-locality ansatz does not prove polynomial sample complexity for this
state-resolved design.  These distinctions are the bridge between the language of
Hamiltonian-learning queries and the settings, shots, and target-gate applications used in QCVV.

The error-generator language also appears in gate-set tomography and device
calibration \cite{blumekohout2017gst,xue2022spin}.  GST's gauge is a self-consistent
basis/SPAM gauge across a gate set; our quotient is an application-level equivalence
group fixed for one diagonal target.  The two should not be conflated.  Gate-set
shadows and other targeted channel estimators can learn broad collections of
functionals \cite{helsen2023gateset}; our deterministic design instead exploits a
known eigenbasis and exposes exactly which Walsh supports each Boolean face sees.

Randomized and cycle benchmarking provide scalable aggregate or cycle-level error
diagnostics \cite{magesan2011rb,erhard2019cycle}, while structured protocols can
learn correlated Pauli noise \cite{harper2020noise}.  Those QCVV objectives are
complementary to, but distinct from, reconstructing the branch-resolved phase field
and testing membership in a declared diagonal quotient.

Which tool to prefer follows from the estimand, not from a precision ranking:
randomized or cycle benchmarking when only an aggregate error rate is needed;
gate-set tomography or self-consistent process characterization when the gate
set and SPAM must be diagnosed jointly; targeted or shadow-style channel
functionals when the register is large and only a few properties matter.  The
present method applies when the successful block is promised diagonal in a
known basis and the question is a declared quotient contract plus the
localization of coherent phase support; these choices are complementary, and
none is unconditionally more scalable or precise.

The central novelty claim is correspondingly combinatorial and contractual, not the
generic phrase ``gate calibration as Hamiltonian learning.''  It consists of the
phase-torus M\"obius criterion, its local Walsh chart, the graph--face support
relation, the multi-target insufficiency result, and a falsifiable route from
generator coordinates to a rank-and-sensitivity-gated control update.

\subsection{How general should a generalized phase gate be?}

There is no target-independent definition of ``generalized.''  For the shared-control
family, $\cG_{\le m}$ is natural because it preserves every intended
$(m+1)$-body control--target factor while allowing all lower-weight phase dressing.
It applies without change to every $C^mZ^{\otimes n}$, but its consequences depend on
$n$ and on external spectators: a single designated face is complete only when it is
the whole register.  For another task, the physically meaningful group may be only
global plus virtual single-qubit $Z$ frames, a hardware/compiler compensation group,
or a different algebra selected by the algorithm.  The pair
$(U_{\rm tar},\cG_{\rm allow})$ must therefore accompany every generalized-gate
number.

The title therefore emphasizes conditional phase gates.  Leakage awareness is
essential to the validity and fidelity layers, but leakage is not a coordinate of
the learned commuting Hamiltonian.  Making it a title modifier would obscure the
primary object; omitting it from the title does not permit it to disappear from
acceptance.

\subsection{Benchmarking and experimental deployment}

The existing Pauli-transfer character benchmarking (PTCB) construction targets a
fixed ideal self-inverse Boolean matrix \cite{ye2026ptcb}.  An uncompensated allowed
dressing can therefore reduce the strict PTCB result even when generalized
membership is perfect.  A generalized target may be compared with that strict
benchmark only after applying and independently validating $V^\dagger$, or after
deriving a new benchmark whose target contract explicitly includes
$\cG_{\rm allow}$.  We do neither here; PTCB remains a proposed held-out strict-target
test, not a signed phase estimator or a quotient-fidelity measurement.

A device implementation would first specify the bit ordering, target unitary, and
allowed and compensable groups.  It would then qualify joint spectator readout and
gate/reference interleaving; establish transfer, leakage, survival, visibility,
branch, loop, and depth thresholds; estimate a baseline generator and a
finite-difference control-response Jacobian; evaluate a local update on training
data; fix the updated parameters; and finally acquire independent phase maps and a
strict held-out benchmark.  Thermal recovery and polarization require separate
qualification.  The experimental inputs are summarized in
\cref{app:deployment}.

\subsection{Limitations and falsification routes}

The main limitations are constructive:
\begin{enumerate}
\item The logarithm is branch-local.  Data on or beyond the capture boundary do not
identify a unique generator without extra continuity or prior information.
\item The diagonal model cannot represent arbitrary off-diagonal coherent error,
leakage dynamics, or non-Markovian variation.  The model-validation tests cover named
alternatives, not every violation.
\item The complete phase map has exponentially many outcomes.  Multiplexed settings
reduce recompilation, not the number of basis phases or rare-bin shot cost.
\item The $4^m$ variance law is local and design-specific, and the
$n2^{n-1+x}$ context factor assumes separate face budgets.  We have not proved a
minimax lower bound against adaptive, entangled, collective, sparse-recovery, or
joint-context protocols.
\item The covariance law in \cref{eq:face-noise-covariance} is per estimation block.
Noise that independently resamples between blocks can be averaged; a quasistatic
offset within the acquisition acts as a floor until an additional drift model or
interleaving scheme is introduced.  A bounded systematic bias does not average at
all.
\item The branch-projection metric is only a deterministic lower bound on the global
quotient optimum away from membership.
\item The primary central-spin update uses the same conditional-block model for the
simulated system and Jacobian.  The response-gain sensitivity study exposes magnitude and sign
errors in one column, but does not demonstrate robustness to general transfer-function
error, dissipation, drift, or an incorrect control model.
\item A low-dimensional generator does not imply that microscopic controls are
identifiable.  The selected-tone-phase negative control is a concrete counterexample.
\item The finite-shot decision of \cref{sec:finite-shot-decisions} is a
local-Gaussian union-bound construction: its nominal familywise error is
asymptotic, $\bm\tau$ and $\alpha$ are prespecified components of the statistical
acceptance contract, and
\textsc{unresolved} is a real third state whose resolution requires more data
or a wider declared tolerance.
\end{enumerate}
Each limitation defines a future falsification experiment rather than a hidden
assumption: enlarge the branch with additional depths, test off-diagonal observables,
compare separate-context and shared-data acquisition, measure temporal covariance,
certify a global quotient optimizer, use a deliberately misspecified validation
simulator, and perform independent Jacobian rank tests.

\section{Conclusion}
\label{sec:conclusion}

Calibration of multiqubit conditional phase gates can be organized as structured
Hamiltonian learning once the learned object is stated correctly.  Ramsey measurements reconstruct a
target-relative phase field; a Walsh transform turns it into a commuting
stroboscopic generator; and a specified target quotient determines which coefficients
are allowed or forbidden.  Boolean target faces isolate intended high-order
conditional phases, but our support theorem shows exactly when they are incomplete.
The resulting separation between targeted-factor correctness, complete generalized
membership, strict matrix fidelity, and leakage-aware projected fidelity prevents a
generalized $CCCZ$ representative from being misclassified while also preventing a
multi-target blind interaction from passing as a complete gate.

For the specified independent, uniform Ramsey design, a face formed from $m$ controls
has variance at least $4^{m+1}/(N_{\rm face}\Gamma)$, while an exhaustive
separately-budgeted audit has $n2^{n-1+x}$ target--context outputs.  Coherence and
survival determine the useful repetition depth through $\Gamma$; context-readout
inversion can amplify variance exponentially; correlated phase noise is selected by
the face-sign geometry; and systematic bin errors can add coherently as $2^m$.
These statements are explicit protocol bounds and noise laws, not a universal
Hamiltonian-learning exponent.

The seeded simulations test the estimator, its model checks, the fixed-design
scaling laws, and one same-model learning-to-control step.  Separate tests verify
the global branch-invariant criterion and the $4M$ joint-binning acquisition from
raw counts; the response-gain sweep makes one control-model failure boundary
explicit.  The negative results are equally important: high-order rare-bin cost is
visible, nonrepeating dynamics are rejected rather than averaged, a target-face
audit can miss a forbidden interaction, and a nearly null or sign-reversed
generator-to-control response does not become a calibration claim.  Experimental
precision, physical transfer, and held-out gate benchmarking remain open.

\acknowledgments
We thank fruitful discussions with Xiongfeng Ma, Guoding Liu, and Yu He for the methodology of this work. This work was supported by the Key-Area Research and Development Program of Guangdong Province
(Grant No. 2018B030326001), the Science, Technology and Innovation Commission of Shenzhen Municipality
(Grant Nos. JCYJ20170412152620376 and KYTDPT20181011104202253), the Innovation Program for
Quantum Science and Technology (2021ZD0301703), Guangdong Major Project of Basic Research
(2025B0303000007), the Shenzhen Science and Technology Program (Grant No. KQTD20200820113010023),
and CCF-QuantumCtek Superconducting Quantum Computing Special Cooperation Program (No.
CCF-QC2025002).

\appendix

\section{Proofs of the structural results}
\label{app:proofs}

\subsection{Walsh orthogonality and the generator}

For $S,R\subseteq[M]$, factorization over Boolean coordinates gives
\begin{equation}
2^{-M}\sum_z\chi_S(z)\chi_R(z)=\delta_{SR}.
\label{eq:walsh-orthogonality}
\end{equation}
Thus the $2^M$ characters form an orthonormal basis for real phase fields and
\begin{equation}
\epsilon(z)=\sum_S\widehat\epsilon_S\chi_S(z).
\end{equation}
Because the $Z_S$ commute,
\begin{equation}
\exp\!\left(i\sum_S\widehat\epsilon_SZ_S\right)\ket z
=\e^{i\epsilon(z)}\ket z.
\end{equation}
Within the injective chart this exponent is the selected logarithm, proving
\cref{thm:generator}.  A global phase $\gamma$ adds
$2^{-M}\sum_z\gamma\chi_S(z)=\gamma\delta_{S,\varnothing}$ and leaves every
nonempty coefficient unchanged.

For \cref{thm:quotient}, the Walsh basis makes
$H_{\rm err}=H_{\rm allow}+H_{\rm forbid}$ a unique orthogonal decomposition.  If
$H_{\rm forbid}=0$, then $H_{\rm err}\in\mathfrak g_{\rm allow}$ and, because its
eigenphase vector lies in $\mathcal C$,
$U_{\rm err}\in\cG_{\rm allow}^{\mathcal C}$.  Conversely, membership in
$\cG_{\rm allow}^{\mathcal C}$ supplies an allowed $H$ whose eigenphase vector lies
in the same injective chart.  Injectivity gives $H=H_{\rm err}$, so the forbidden
projection vanishes.  The chart restriction is indispensable: state-dependent
multiples of $2\pi$ can change Walsh coefficients without changing the unitary.

\subsection{Multiplicative M\"obius membership}

For an arbitrary map $u:\{0,1\}^M\to U(1)$, let
$g_T=\Mob_T[u]$.  M\"obius inversion in the multiplicative Abelian group $U(1)$
gives the exact identity
\begin{equation}
u(z)=\prod_{T\subseteq\operatorname{supp}(z)}g_T.
\label{eq:multiplicative-inversion}
\end{equation}
If $u(z)=\exp[ip(z)]$ for a real Boolean polynomial of degree at most $m$, every
finite difference above degree $m$ vanishes modulo $2\pi$; hence $g_T=1$ for
$|T|\ge m+1$.  Conversely, suppose all those $g_T$ equal one.  Choose any real
$\beta_T$ satisfying $\e^{i\beta_T}=g_T$ for $|T|\le m$ and define
\begin{equation}
p(z)=\sum_{\substack{T\subseteq\operatorname{supp}(z)\\|T|\le m}}\beta_T.
\end{equation}
Equation~\eqref{eq:multiplicative-inversion} then gives $u(z)=\e^{ip(z)}$.  This is
a Boolean polynomial in number projectors of degree at most $m$; because
$n_j=(\Id-Z_j)/2$, it lies in the same diagonal operator space as Pauli-$Z$ strings
of weight at most $m$.  Thus the relative unitary belongs to $\cG_{\le m}$, proving
\cref{thm:global-membership}.  No phase representative or logarithm was selected in
either direction.

\subsection{Face--Walsh identity}

Apply $D_A$ to one character.  If some $i\in A$ is absent from $S$, the two terms
differing only in $x_i$ cancel.  If $A\subseteq S$, each active coordinate contributes
$\chi_i(1)-\chi_i(0)=-2$, while complementary coordinates give
$\chi_{S\setminus A}(r)$.  Therefore
\begin{equation}
D_A\chi_S(r)=
\begin{cases}
0,&A\not\subseteq S,\\
(-1)^{|A|}2^{|A|}\chi_{S\setminus A}(r),&A\subseteq S.
\end{cases}
\end{equation}
Linearity proves \cref{eq:face-walsh}.  A second Walsh transform over the
complementary word yields
\begin{equation}
2^{-|R|}\sum_r\chi_B(r)D_A\epsilon(r)
=(-1)^{|A|}2^{|A|}\widehat\epsilon_{A\cup B},
\label{eq:spectator-walsh}
\end{equation}
which proves exact recovery of every superset coefficient.

\subsection{Completeness and the blind support}

For $n=1$ with no external spectators, the register size is $M=m+1$ and
$A_1=[M]$.  The only support with $|S|\ge m+1$ is therefore $S=A_1$, so the target
face spans the forbidden sector.

For $n\ge2$, $S_*$ in \cref{eq:blind-support} has
$|S_*|=(m-1)+2=m+1$, hence is forbidden.  It lacks $c_*$, so neither
$C\cup\{t_1\}$ nor $C\cup\{t_2\}$ is a subset; it also omits every other target.
The face identity then gives $D_{A_j}\chi_{S_*}=0$ for all $j$, proving
target-face blindness.  On the other hand, the $S_*$ M\"obius invariant of
$u(z)=\exp[ia\chi_{S_*}(z)]$ is
$\exp[i(-2)^{m+1}a]$.  It is nontrivial precisely when
$2^{m+1}a\notin2\pi\mathbb Z$, completing the proof of
\cref{prop:blind-support}.  Replacing $t_2$ by an external spectator proves the
single-target-with-spectator statement.

\subsection{Graph identifiability}

If $Bx=0$, endpoint values of $x$ agree across every measured edge.  Connectivity
propagates equality along paths, so $\ker B$ contains only constants and
$\rank B=|V|-1$.  Removing one reference coordinate gives full column rank and the
unique GLS solution \cref{eq:graph-gls}.  A linear functional is constant on the
solution class $\bm h+\ker X$ exactly when it annihilates $\ker X$.  The fundamental
theorem of linear algebra, $(\ker X)^\perp=\operatorname{row}X$, proves
\cref{eq:row-space-test}.

\section{Complexity and noise derivations}
\label{app:complexity-proofs}

\subsection{Arbitrary-contrast bound and allocation}

Order the independent local phase parameters by state $z$ and arm $a\in\{g,r\}$.
Their Fisher matrix is diagonal with entries $I_z^{(a)}$.  The gradient of
$\vartheta_q=\sum_zq_z(\phi_z^{(g)}-\phi_z^{(r)})$ has entries $q_z$ in the gate
arm and $-q_z$ in the reference arm.  The scalar Cram\'er--Rao inequality therefore
gives \cref{eq:general-contrast-crb}.  Writing $I_z^{(a)}=N_z^{(a)}\Gamma_z^{(a)}$
reduces shot allocation to minimizing
\begin{equation}
\sum_{z,a}\frac{q_z^2}{N_z^{(a)}\Gamma_z^{(a)}}
\quad\text{subject to}\quad \sum_{z,a}N_z^{(a)}=N.
\end{equation}
Cauchy--Schwarz yields
\begin{equation}
\left(\sum_{z,a}\frac{q_z^2}{N_z^{(a)}\Gamma_z^{(a)}}\right)N
\ge
\left(\sum_{z,a}\frac{|q_z|}{\sqrt{\Gamma_z^{(a)}}}\right)^2,
\end{equation}
with equality exactly when
$N_z^{(a)}\propto|q_z|/\sqrt{\Gamma_z^{(a)}}$.  This proves
\cref{eq:general-optimal-allocation}.

\subsection{Proof of the fixed-design face bound}

For one accepted Bernoulli observation,
$p=(1+V\cos\vartheta)/2$,
$\partial_\delta p=-qV\sin\vartheta/2$, and
$p(1-p)=(1-V^2\cos^2\vartheta)/4$.  The accepted-shot Fisher information is
\begin{equation}
\frac{(\partial_\delta p)^2}{p(1-p)}
=\frac{q^2V^2\sin^2\vartheta}
{1-V^2\cos^2\vartheta}\le q^2V^2,
\end{equation}
with equality at $\cos\vartheta=0$.  If successful-block heralding is independent
of $\delta$, a rejected attempt has zero phase score and an accepted attempt occurs
with probability $s$.  Averaging the conditional score therefore multiplies the
information by $s$.  Independent attempts and depths add, proving
\cref{eq:bin-information}.

Order the parameter vector as all gate-bin phases followed by all reference-bin
phases.  Independence makes its Fisher matrix diagonal, with diagonal entries
$I_c^{(g)}$ and $I_c^{(r)}$.  The gradient of
$\delta_A=\sum_ca_c(\delta_c^{(g)}-\delta_c^{(r)})$ has components
$a_c$ and $-a_c$, where $a_c^2=1$.  The scalar Cram\'er--Rao inequality gives
\[
\Var(\widehat\delta_A)\ge
\nabla\delta_A^T\mathcal I^{-1}\nabla\delta_A
=\sum_c[(I_c^{(g)})^{-1}+(I_c^{(r)})^{-1}],
\]
which proves \cref{thm:face-crb}.

For $b=2^m$ uniform words, each arm/bin receives
$N_{\rm face}/(2b)$ attempts.  Substitution gives
$b[2b/(N_{\rm face}\Gamma)+2b/(N_{\rm face}\Gamma)]
=4b^2/(N_{\rm face}\Gamma)$, proving \cref{eq:rare-bin-law}.
For unequal arm informations, minimize
$b^2/(N_g\Gamma_g)+b^2/(N_r\Gamma_r)$ subject to
$N_g+N_r=N_{\rm face}$.  The stationary condition is
$N_g/N_r=\sqrt{\Gamma_r/\Gamma_g}$; strict convexity makes it the unique minimum
and gives \cref{eq:optimal-arm-allocation}.

\subsection{Context, confidence, and coordinate counts}

For target $j$, the complementary word contains $n-1$ other targets and $x$
external spectators.  It therefore has $2^{n-1+x}$ assignments.  Summing over
$j=1,\ldots,n$ proves \cref{eq:face-family-size}.  Under separate budgets and a
Gaussian local limit, a two-sided Bonferroni threshold is
$z_{1-\alpha/(2L)}$.  Power $1-\beta$ against magnitude $\Delta$ follows from
$\Delta/\mathrm{SE}\ge z_{1-\alpha/(2L)}+z_{1-\beta}$.  Substitution of
\cref{eq:rare-bin-law} proves \cref{eq:simultaneous-detection}.

A Hermiticity-preserving linear map on $d\times d$ matrices has $d^4$ real
coordinates, and trace preservation imposes $d^2$ independent real constraints in
the interior, giving $d^4-d^2$.  A unitary has $d^2$ real coordinates and one
channel-invisible global phase.  A diagonal unitary has $d$ phases and the same
one-dimensional gauge.  Counting nonempty Pauli-$Z$ supports through weight $k$
gives $\sum_{\ell=1}^k\binom M\ell$; counting supports above weight $m$ gives
$\sum_{\ell=m+1}^M\binom M\ell$.  These identities establish the coordinate rows
of \cref{tab:complexity}, without converting them into shot bounds.

\subsection{Depth, confusion, covariance, and bias}

Taking the logarithm of \cref{eq:markov-information} gives
$2\log q-aq+\mathrm{const}$ with $a=2\gamma+\ell$.  Its derivative vanishes at
$q=2/a$.  Dividing by $q$ for a target-gate-call budget changes the derivative to
$1/q-a$ and the optimum to $1/a$.  For an offset
$\xi\sim\mathcal N(0,\sigma^2)$ that is static within a circuit and independently
redrawn between attempts,
$\E\e^{iq\xi}=\e^{-\sigma^2q^2/2}$, so the information contains
$\e^{-\sigma^2q^2}$.  Differentiating then gives $1/\sigma$ per attempt and
$1/(\sqrt2\,\sigma)$ per target-gate call when $a=0$.

One symmetric context bit has confusion matrix
$C_1=\bigl(\begin{smallmatrix}1-r_c&r_c\\r_c&1-r_c\end{smallmatrix}\bigr)$
and singular values $1$ and $1-2r_c$.  For independent bits,
$C_h=C_1^{\otimes h}$; singular values of a tensor product multiply.  Hence
$\sigma_{\min}(C_h)=(1-2r_c)^h$ and
$\|C_h^{-1}\|_2^2=(1-2r_c)^{-2h}$, proving
\cref{eq:readout-amplification}.

Equation~\eqref{eq:face-noise-covariance} is ordinary linear covariance
propagation.  Since a nonempty Boolean face has $\bm a^T\bm1=0$ and
$\|\bm a\|_2^2=2^m$, it gives zero for
$C_\phi=\sigma^2\bm1\bm1^T$, $2^m\sigma^2$ for
$C_\phi=\sigma^2I$, and $4^m\sigma^2$ for
$C_\phi=\sigma^2\bm a\bm a^T$.  The triangle inequality gives
$|\sum_ca_cb_c|\le\sum_c|b_c|\le2^mb_{\max}$; choosing
$b_c=a_cb_{\max}$ proves tightness.  Finally, the weighted through-origin slope
for $y_q=q\delta+\beta_q$ gives the first expression in \cref{eq:depth-bias};
weighted Cauchy--Schwarz gives the second.

\subsection{Simultaneous finite-shot decision}

For the local-Gaussian regime, $\widehat{\boldsymbol\mu}$ is asymptotically
normal with covariance $C_\mu$ from \cref{eq:mu-covariance}; under the null
each studentized coordinate is approximately standard normal.  The Bonferroni
union bound therefore gives, for any dependence structure,
\begin{align}
\Pr\!\left(\exists T:\ |Z_T|>c_\alpha\right)
&\le\sum_{T\in\mathcal H_m}\Pr(|Z_T|>c_\alpha)\nonumber\\
&\le2L_{\rm global}[1-\Phi(c_\alpha)]=\alpha,
\label{eq:fw-bound}
\end{align}
by the choice in \cref{eq:simultaneous-critical}.  Because
$|\widehat\mu_T|+c_\alpha\sigma_T\le\tau_T$ implies the entire interval lies
inside the tolerance band, a \textsc{pass} certifies every forbidden
coordinate within $\bm\tau$ at simultaneous level at least $1-\alpha$, and a
\textsc{fail} identifies a coordinate whose interval excludes tolerance.  The
$\arg\Mob_T$ observables are unchanged by state-dependent $2\pi$ relabelings,
so the decision is branch independent, while the linearization remains valid
only inside the accepted chart.

\section{Estimator and uncertainty details}
\label{app:statistics}

\subsection{Fisher calculation}

The exact information and the face Cram\'er--Rao proof are in
\cref{app:complexity-proofs}.  The finite-sample estimator retains the actual
gate/reference, depth, survival, and bin-level resource accounts rather than
replacing them by the uniform corollary.  Local curvature is compared with the
corresponding finite-sample prediction.

The finite-sample implementation refuses a face estimate unless every required bin
passes its minimum-count rule.  Conditional covariance is reported separately from
the probability that an attempted circuit passes this criterion.  This distinction
was necessary in exploration:
under rare-bin stress, nominal covariance could look accurate among retained runs
while most runs were invalid.

\subsection{Circular fit and residual}

For each state or edge, the implementation forms the joint binomial log likelihood
of gate/reference $X$ and $Y$ counts at every declared depth.  It searches only the
prespecified phase-estimation window around the prediction.  Local curvature supplies
a conditional standard error.  A competing local minimum, a boundary minimum, or a
phase field outside the accepted chart invalidates the estimate.  A Pearson-like
depth residual is compared with a prespecified threshold; it is not used to enlarge error
bars after seeing a discrepancy.

\subsection{Meaning of the plotted error bars}

In \cref{fig:global-validation}(a), the plotted curve is a deterministic alias
regression and has no uncertainty bars.  In panel (b), each point is the arithmetic
mean of 200 per-map RMSE values and each bar is their ordinary sample SD, not the
uncertainty of the mean.  In panel (c), each height is a rejection count divided by
300 and each interval is the score-Wilson interval in \cref{eq:wilson}; the four
intervals are recomputed independently from their own indicators.

In \cref{fig:generic-validation}(a,c), each plotted point is $\widehat p=k/N$ for
independent Monte Carlo rejection indicators.  The displayed score-Wilson interval
is
\begin{equation}
\frac{\widehat p+z^2/(2N)}{1+z^2/N}
\ \pm\ 
\frac{z}{1+z^2/N}
\sqrt{\frac{\widehat p(1-\widehat p)}{N}+\frac{z^2}{4N^2}},
\label{eq:wilson}
\end{equation}
with $z=1.9599639845$.  It remains nonzero at $k=0$ or $k=N$ and therefore avoids
the misleading zero-width Wald bar.  The intervals are asymmetric; the vector figure
draws their exact lower and upper endpoints.  Panel (b) reports RMSE without an
uncertainty bar because no bootstrap or analytic RMSE interval was prespecified.

In \cref{fig:complexity-noise}, all points are sample variances recomputed from the
saved replicate arrays.  Solid lines are analytic predictions.  The two dashed
lines in panel (a) are the prespecified ratio criterion, not a confidence or
credible interval.  No per-point error bar is shown because no variance-estimator
interval was specified before the confirmatory run.  The absence of such bars is
therefore stated explicitly.

In \cref{fig:control}(a,b), the baseline is a deterministic simulator value and the
post-update metric varies because the training counts vary.  Error bars are the
ordinary sample SD of the 30 post-update outcomes; they are not the SEM and are not a
shared visual constant.  Panel (c) contains deterministic finite-difference
sensitivities and has no uncertainty bars.  Each panel therefore states the
statistical meaning of its plotted summaries explicitly.

\section{Fidelity identities}
\label{app:fidelity}

The Haar second moment satisfies
\begin{equation}
\E_\psi[(\proj\psi)^{\otimes2}]=
\frac{\Id+\mathrm{SWAP}}{d(d+1)}.
\end{equation}
For one successful-block operator $L$, contraction gives
\begin{equation}
\E_\psi|\langle\psi|L|\psi\rangle|^2=
\frac{\Tr(L^\dagger L)+|\Tr L|^2}{d(d+1)}.
\end{equation}
Taking $L=U_{\rm tar}^\dagger K$ proves \cref{eq:projected-fidelity}; unitary $L$
proves \cref{eq:strict-fidelity}.  Average fidelity is one exactly when all
eigenvalues of the relative unitary coincide, so $F_{\rm gen}=1$ is equivalent to
quotient membership.

For the factorized error model, all $(2^m-1)2^n$ words whose controls are not all one
have relative phase zero.  On the all-one control word, summing over target bits
factorizes:
\begin{equation}
\sum_{t\in\{0,1\}^n}\e^{i\sum_j\delta_jt_j}
=\prod_j(1+\e^{i\delta_j}),
\end{equation}
which proves \cref{eq:factor-trace}.

\section{Requirements for experimental deployment}
\label{app:deployment}

Applying the protocol to a device requires an explicit computational-bit ordering
and target word, a definition of reset within the gate cycle, and a declaration of
the virtual-$Z$ and multiqubit phase operations that are actually compensable.  The
measurement model must include the calibrated joint-readout confusion matrix,
leakage-resolved outcomes, gate/reference timing, and relevant thermal or
polarization recovery.  Control bounds, branch and loop thresholds, and the
allocation for an independent strict-target benchmark must be fixed before the
corresponding data are analyzed.

To separate calibration from validation, the experimental sequence should be
\begin{equation}
\boxed{\begin{gathered}
\text{qualify model}\ \to\ \text{estimate generator}\\
\to\ \text{estimate response}\ \to\ \text{update on training data}\\
\to\ \text{fix controls}\ \to\ \text{independent test}
\end{gathered}}
\end{equation}
Validation data used to support a final performance statement cannot also be used
to choose the reported update.  A subsequent update requires a new independent
validation data set.  The numerical studies in this article exercise the learning
logic and selected failure modes, but no step in this deployment sequence has been
performed on a device here.

\end{document}